\documentclass[manuscript]{aastex}
\usepackage{graphicx}
\usepackage{color}

\definecolor{red}{rgb}{0.0,0.0,0.0}

\slugcomment{Not to appear in Nonlearned J., 45.}

\shorttitle{Artifacts and PSFs of the IRC}
\shortauthors{Ko Gakkachou Arimatsu et al.}

\begin{document}


\title{Characterization and Improvement of the Image Quality of the Data Taken with the Infrared Camera (IRC) Mid-Infrared Channels onboard {\it AKARI}}


\author{Ko Arimatsu\altaffilmark{1}, Takashi Onaka\altaffilmark{1},
Itsuki Sakon\altaffilmark{1}, Shinki Oyabu\altaffilmark{2}, 
Yoshifusa Ita\altaffilmark{3}, Toshihiko Tanab\'{e}\altaffilmark{4}, Daisuke Kato\altaffilmark{5}, Fumi Egusa\altaffilmark{5},
Takehiko Wada\altaffilmark{5}, and Hideo Matsuhara\altaffilmark{5}}
%\affil{Astronomy Department, University of California,
%    Berkeley, CA 94720}

%\affil{National Optical Astronomy Observatories, Tucson, AZ 85719}
%\email{aastex-help@aas.org}

%\and

%\author{R. J. Hanisch\altaffilmark{5}}
%\affil{Space Telescope Science Institute, Baltimore, MD 21218}


\altaffiltext{1}{Department of Astronomy, Graduate School of Science, The University of Tokyo,
7-3-1 Hongo, Bunkyo-ku, Tokyo 113-0033, Japan.}
\email{arimatsu@astron.s.u-tokyo.ac.jp}
\altaffiltext{2}{Graduate School of science, Nagoya University, Chikusa-ku, Nagoya 464-8602, Japan.}
%\altaffiltext{2}{Okayama Astrophysical Observatory, NationalAstronomical Observatory, Kamogata, Asakuchi, Okayama 719-0232, Japan.}
%\altaffiltext{3}{Department of Physics and Astronomy, University of Denver, 2112 E. Wesley Ave., Denver, CO 80208, USA.}
\altaffiltext{3}{Astronomical Institute, Graduate School of Science, Tohoku University, 6-3 Aramaki Aoba, Aoba-ku, Sendai, Miyagi 980-8578, Japan}
\altaffiltext{4}{Institute of Astronomy, School of Science, The University of Tokyo, Mitaka, Tokyo 181-0015, Japan}
\altaffiltext{5}{Institute of Space and Aeronautical Science, Japan Aerospace Exploration Agency, 3-1-1 Yoshino-dai, Chuo-ku, Sagamihara, Kanagawa 252-5210, Japan.}
%%\altaffiltext{4}{Visiting Programmer, Space Telescope Science Institute}
%%\altaffiltext{5}{Patron, Alonso's Bar and Grill}


\begin{abstract}
Mid-infrared images 
%obtained with the Infrared Camera (IRC) onboard {\it AKARI}
\textcolor{red}{frequently} 
suffer artifacts and extended point spread functions (PSFs).
%when a bright object is present in the field-of-view.
We investigate the characteristics of the artifacts and the PSFs
 in images obtained with 
%mid-infrared channels 
\textcolor{red}{the Infrared Camera (IRC) onboard {\it AKARI}}
at 
\textcolor{red}{four mid-infrared bands of} 
the S7 (7\,$\mu$m), S11 (11\,$\mu$m), L15 (15\,$\mu$m), and L24 (24\,$\mu$m).
% bands.
Removal of the artifacts significantly improves the reliability of the reference data for flat-fielding at the L15
and L24 bands.
%With artifact models for MIR-L channel, we revise flat frames without ghost effects for imaging data of L15 and L24.
A set of models of the IRC PSFs is also constructed from on-orbit data.
These PSFs have extended components that come from diffraction and scattering within the detector arrays.
We estimate the aperture correction factors for point sources 
and the surface brightness correction factors for diffuse sources.
We conclude that the surface brightness correction factors 
range from  0.95 to 0.8, taking account of the extended component of the PSFs.
To correct for the extended PSF effects for the study of faint structures, 
we also develop an image reconstruction method, 
which consists of the deconvolution with the PSF and the convolution with an appropriate Gaussian.
The appropriate removal of the artifacts, 
improved flat-fielding,
 and image reconstruction with the extended PSFs enable us to investigate detailed structures of extended sources in IRC mid-infrared images.
%\textcolor{red}{The present results provide a general method to improve the quality of the IRC mid-infrared data.  They will also be useful for the design and data analysis for future mid-infrared instruments.}
%%This PSF correction method offers the diffuse absolute calibration of diffuse objects and to correct the contamination from nearby bright objects.
\end{abstract}



\section{Introduction}
\textcolor{red}{Mid-infrared multiband imaging data 
offer a useful tool to study dusty stellar objects 
(e.g., \citealt{meixner08,ita08}) 
as well as the interstellar medium (e.g., \citealt{bernard08,onaka10}).  
However, mid-infrared observations frequently suffer 
from various
instrumental effects, 
such as the anomalies in the detector performance \citep{pipher04},
the extended point spread functions (PSFs) due to the diffraction of light \citep{Jarrett06}, 
and artifacts.
%%%%%
The corrections for these annoying effects should become crucial particularly 
when we are interested in the faint diffuse objects adjacent to bright objects.
In this {\it Paper} we report detailed investigations on the qualities of the
mid-infrared imaging data taken with the Infrared Camera on board Japanese {\it AKARI} satellite 
to evaluate their artifacts and to offer useful information needed for the advanced data 
reduction procedures to users of the {\it AKARI} mid-infrared data, 
which are now publically available from the data archive server.}
% extended PSFs, and adjustment of the PSFs.

%which have to be properly corrected, in particular for the study of diffuse emission.}  
%The instrumental effects of mid-infrared instruments inherently originates 
%in anomalies in the detector performance (e.g., Pipher et al. 2004) and the diffraction of light, 
%which produce extended point spread functions (PSFs; \citealt{Jarrett06}).  
%This makes calibration for the diffuse emission difficult (\citealt{reach05,cohen07}).  
%Part of the effects also come from large refractive indexes of optical materials available for mid-infrared ($> 3$; e.g., Si, ZnSe, or KRS-5), for which perfect anti-refection (AR) coating is difficult in a wide spectral range.  
%Imperfect AR coating often produces non-negligible artifacts in mid-infrared images (e.g. \citealt{okumura03}). 
%These effects significantly degrade the quality of MIR images and hamper the detailed study of extended structures of an object.
% Furthermore, a wide spectral range of mid-infrared observations requires proper adjustment of their PSFs of different wavelength band 
%(e.g. \citealt{gordon08}).  
%In this {\it Paper} we report detailed investigations on mid-infrared data taken with the Infrared Camera on board Japanese AKARI satellite to investigate their artifacts, extended PSFs, and adjustment of the PSFs.}  

The infrared satellite {\it AKARI} carried out about 5000 pointed observations 
in addition to the all sky survey at the mid- to far-infrared bands during its cold phase (until 2008 Aug 24) \citep{murakami07}.
 The Infrared Camera (IRC: \citealt{onaka07}) on board {\it AKARI} made 
near- to mid-infrared (2--26\,$\mu$m) imaging as well as spectroscopic observations in the pointed observation mode.
The IRC has three channels: NIR, MIR-S and MIR-L,
which cover the spectral ranges 1.8--5.5\,$\mu$m, 4.6--13.4\,$\mu$m, and 12.6--26.5\,$\mu$m, respectively.
Each channel has a field-of-view of about $10\arcmin \times 10\arcmin$ and is equipped with three \textcolor{red}{medium-band} filters \textcolor{red}{for imaging observations
(see \citealt{onaka07} for details)}.
%For details of the IRC design, see \citet{onaka07}.

%\textcolor{red}{Here} we investigate the imaging performance of the two mid-infrared (MIR) channels of the IRC, MIR-S and MIR-L.
%In particular, we investigate the artifacts and the extended point spread functions (PSFs)
%at four bands (S7 (7\,$\mu$m), S11 (11\,$\mu$m), L15 (15\,$\mu$m), and L24 (24\,$\mu$m)).
%Hereafter, images taken with these four bands are called MIR images.  
When a bright object comes into the field-of-view, 
several artifacts, 
\textcolor{red}{often termed} as "ghosts" (e.g. \citealt{lorente08}),
 appear in MIR images 
\textcolor{red}{of S7 (7\,$\mu$m), S11 (11\,$\mu$m), L15 (15\,$\mu$m), and L24 (24\,$\mu$m) bands}
due to reflections among the optical elements \citep{lorente08}.
The scattering of light internal to the Si:As detector arrays 
\textcolor{red}{of the {\it AKARI}}
also spreads out the signals from the object over the entire array. 
This scattering together with the diffraction produces very extended components in the PSFs.
It is not straightforward to separate artifacts from the PSF unambiguously.
In this {\it Paper}, 
we define artifacts as components that move differently from the 
real object in the image or those showing very distinct, asymmetric 
patterns (\S~\ref{artifact}).
%we define artifacts as components that move
%differently from real objects in the image (\S~\ref{sub_Lartifact}) or those showing very
%distinct, asymmetric patterns (\S~\ref{sub_Sartifact}). 
\textcolor{red}{In our definition, 
anomalies of the detector array and the effects of diffraction are included in the PSF.
We present high dynamic range (HDR) PSFs, 
aperture correction factors, 
and surface brightness correction factors for diffuse sources at S7, S11, L15, and L24 bands
(\S~\ref{psf}).
The obtained HDR PSFs are useful in the process of image reconstruction (see \S~\ref{IRM}),
which enable us not just to carry out an accurate photometry of point sources 
but also to get much improved information on the structures and surface brightness of diffuse extended sources.
We also obtain the revised reference data for flat-fielding 
(hereafter flat data) for L15 and L24 that free from the effects of the artifacts (see \S~\ref{flat}).}

\textcolor{red}{Among the information obtained in this study, 
the information commonly used in the process of data reduction (e.g., new flat dataset for L15 and L24) 
is included in the latest update of the imaging toolkit ver. 20110225. 
On the other hands, further advanced data reduction procedures, 
beyond the treatments made with the standard pipeline, 
may be necessary for the data of relatively faint extended sources 
and for those affected by artifacts due to an inclusion of bright objects in the FOV. 
This paper aims to provide the information needed in the course of such advanced data reduction procedures 
to the general users of the {\it AKARI} MIR datasets and 
to demonstrate the method of imaging reconstruction and its application 
to the {\it AKARI} MIR data of M81 to exemplify the usage of information newly given in this paper.}

\section{Detailed examination on the image qualities of {\it AKARI}/IRC data}

\subsection{Artifacts in IRC images \label{artifact}}
\subsubsection{MIR-S artifacts \label{sub_Sartifact}}
\begin{figure}[!ht]
\begin{center}
\includegraphics[width=\hsize,angle=0]{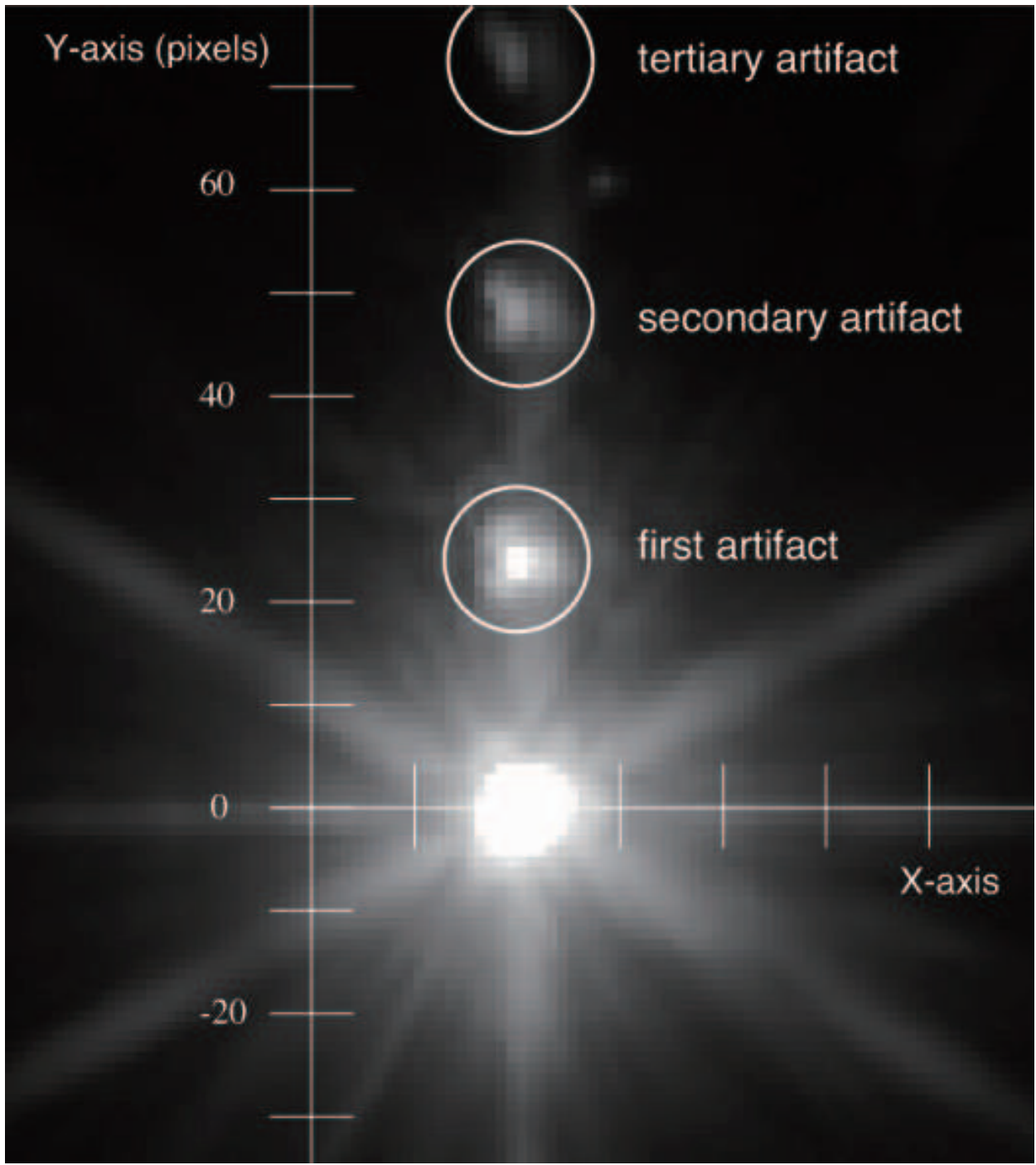}
\caption{Example of the MIR-S artifact at S11 for CRL 618 (pointing ID:4080002.1). 
The secondary and tertiary artifacts, which come from the first and secondary artifacts, are also seen.
The image is displayed in a logarithmic scale.
\label{fig_s0}}
\end{center}
\end{figure}

In the MIR-S channel, a point-like artifact appears with an interval of about 24 pixels from the corresponding real point source in the scan direction (hereafter, defined as the Y-axis), which is attributed to internal reflections in the beam splitter 
\textcolor{red}{since it does not split the light perfectly} (Figure~\ref{fig_s0}).
We employ  archival data of CRL618 and Red Rectangle Nebula of Open Time Programs as a calibration source for the artifact, 
\textcolor{red}{which are listed in Table~\ref{tab4}.}
The outskirts of the PSF are fairly symmetric about the central source and the artifact shows a distinct, point-like pattern.
Using this characteristic, we derive the artifact patterns at the S7 and S11 bands 
by subtracting the signals 24 pixels away from the real source in the opposite side from the image as the background.
No significant difference is noticed in the spatial pattern of the artifact between the two objects.

\begin{figure}[!ht]
\begin{center}
\includegraphics[width=100mm,angle=0]{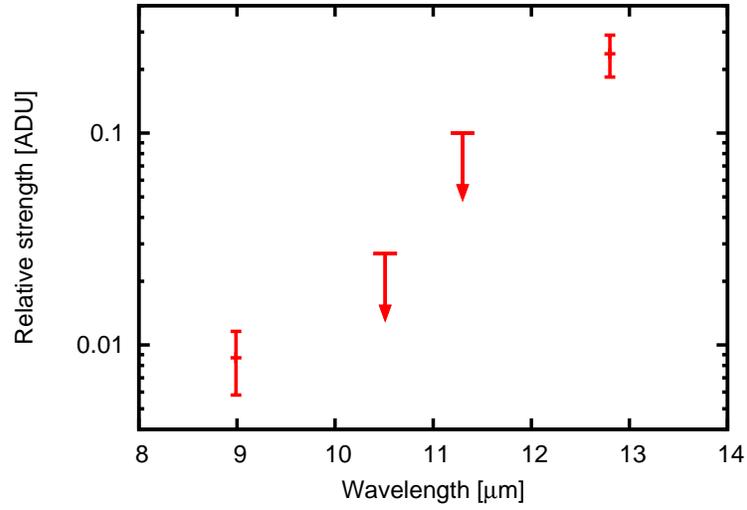}
\caption{Spectral dependence of the artifact in the MIR-S images
derived from emission lines in the slit spectroscopy.
At 10.5 ([\ion{S}{4}]) and 11.3 (UIR)\,$\mu$m the corresponding artifacts are not detected and 3-$\sigma$ upper limits are shown.}
\label{fig.sed}
\end{center}
\end{figure}

Despite the consistency in the spatial pattern of the artifact, 
the ratio of the intensity of the artifact to that of the original object varies among objects.
\textcolor{red}{Since the degree of imperfection varies with wavelength, the intensity of the artifacts can have strong spectral dependence (cf. \citealt{okumura03})}.
We investigate the wavelength dependence of the artifact by using spectroscopic datasets.
The IRC has a narrow slit (Ns) for spectroscopy of diffuse emission \citep{onaka07,ohyama07}.
The dispersion direction is along the Y-axis and the artifact of an emission line appears at a different wavelength.
Three datasets of spectroscopic observations of two Galactic \ion{H}{2} regions and a planetary nebula 
are employed to investigate the spectral dependence (see Table~\ref{tab4}). 
We use the lines of [\ion{Ar}{3}] 8.99\,$\mu$m, [\ion{S}{4}] 10.51\,$\mu$m, [\ion{Ne}{2}] 12.81\,$\mu$m, 
and the 
\textcolor{red}{unidentified infrared (UIR)} 
band at 11.3\,$\mu$m.
The relative intensity of the artifact varies with the wavelength of the incident photon from 1 \% to 4.5 \% for 9 to 13\,$\mu$m (see Figure~\ref{fig.sed}).
The tendency of the wavelength dependence is roughly consistent 
with the wavelength dependence of transmission of the beam splitter  (Figure 5 of \citealt{onaka07}).
Because of this characteristic, the intensity of the artifact depends on the spectrum of the object and cannot be estimated straightforwardly. 
In practice, the estimate of the intensity of this artifact has to be made interactively.
Examples of the relative intensity of the artifacts for various objects  are given in Table~\ref{aa}.
\textcolor{red}{The contributions of the artifacts to the total input signals are measured to be less than 1 \% at S7 and about 1-3 \% at S11, the latter of which is consistent with the values obtained from the spectroscopic data.}

\subsubsection{MIR-L artifacts \label{sub_Lartifact}}
\begin{figure}[!ht]
\begin{center}
\includegraphics[width=\hsize,angle=0]{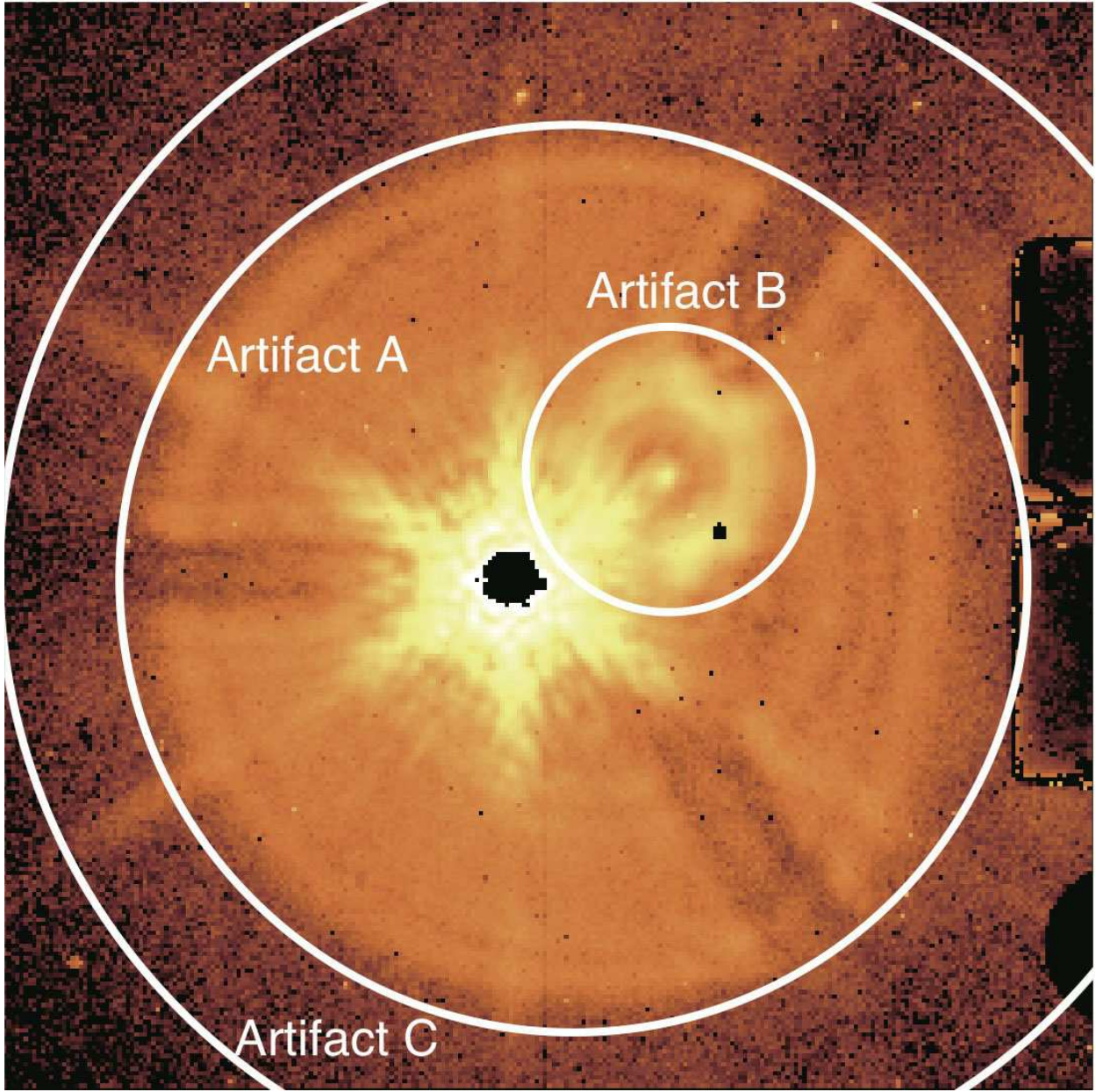}
\caption{
Artifacts at the L24 band. 
The target object is IRAS 22396$-$4708. 
The identified artifacts (A, B, and C) are indicated.
The image is displayed in a logarithmic scale.
The entire field-of-view (256$\times$256 pixels) is shown.
The black regions in the center are saturated pixels. The structures seen in the right edge are the slit mask.
\textcolor{red}{The diffraction spikes and banding (see \S~\ref{psf}) can also be seen in the figure and are overlapped on the artifacts, which can be separated from the artifacts with the separation procedure (see \S~\ref{sub_Lartifact}) and are finally included in the PSF.}\label{fig00}}
\end{center}
\end{figure}

MIR-L images are severely affected by artifacts originating \textcolor{red}{in} reflections among the optical elements and the surface of the detector array
\textcolor{red}{because of the imperfect AR coating}.
There are at least three types of artifacts (A,B and C) 
overlapped with the extended PSF (\S 4) in MIR-L images 
(Figure~\ref{fig00}), 
one relatively compact (artifact B) and the others quite extended (artifacts A and C).
The optical path of each artifact is identified by ray-tracing calculations.
Their major paths are the reflection between the surface of the detector array and the KRS-5 lenses, 
whose 
\textcolor{red}{AR} 
coating \textcolor{red}{is} not perfect particularly at the longer wavelength edge of the spectral range of the MIR-L.
Thus the artifacts appear more strongly at L24 than at L15.
The artifact images from a point source are not focused onto a point, but appear as a diffuse pattern. 
We examine the properties of these artifacts and derive the surface brightnesses and the positions relative to the original object using the on-orbit data listed in Table~\ref{tab4}.
Artifacts A and C, and the PSFs (see \S~\ref{sub_psf}) are very extended and the subtraction of the background sky needs to be made with high accuracy.
There is no appropriate position for the estimate of the background sky in the image,
and we use the data taken at a near sky position that do not have bright objects in their field-of-view as the sky background.

%In one pointed observation, the position of the target object is changed at least three times to carry out a dithering operation.
\textcolor{red}{During an pointed observation carried out with AOT IRC02 \citep{onaka07}, a dithering operation was carried out to settle the target at three different positions on the detector array.}
Using these dithered data, the movement of the artifacts is estimated by a least-square fit for the region where each component of the artifacts does not overlap with each other.
The relative movement of each artifact is assumed
 to be proportional to the movement of the original object. 
Thus the relative movement of the reference position of each artifact $(x_{r}, y_{r})$ from the origin $(x_{o},y_{o})$ is given by 
the position of the incident object $(x_i,y_i)$ as
\begin{eqnarray}
(x_{r}-x_{o}, y_{r}-y_{o}) = (a (x_i-x_{o}),b (y_i-y_{o})),
\label{eqart}
\end{eqnarray}
where $a$ and $b$ are the fitting parameters. 
We set the origin as the center of the array, $(x_{o}, y_{o}) = (128 \,{\rm pix}, 128 \,{\rm pix})$. 
The artifact patterns do not change with the position of the object in the field-of-view 
and thus the reference position $(x_{r}, y_{r})$ is defined as a position near the center of the artifact.
For real objects 
\textcolor{red}{and their PSFs in the image}, 
$a$ and $b$ should be unity, 
but these are not for the artifacts.
By the difference in the relative movement, the artifacts can be separated from the PSF pattern.
The derived fitting parameters are listed in Table~\ref{tab:03}.

\begin{figure}[!ht]
%\begin{center}
   \begin{tabular}{ccc}
      \resizebox{50mm}{!}{\includegraphics[angle=0]{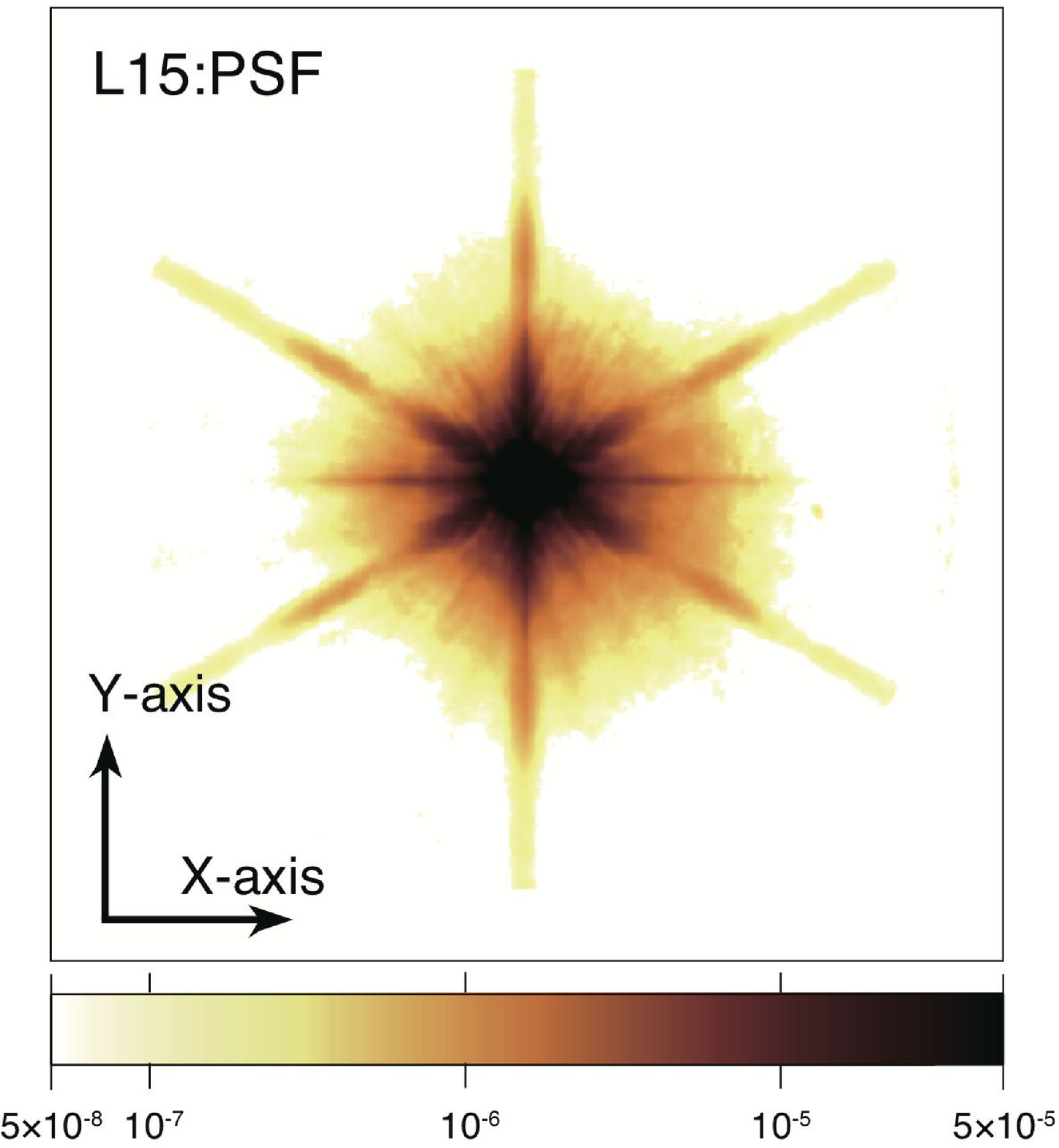}} &
      \resizebox{50mm}{!}{\includegraphics[angle=0]{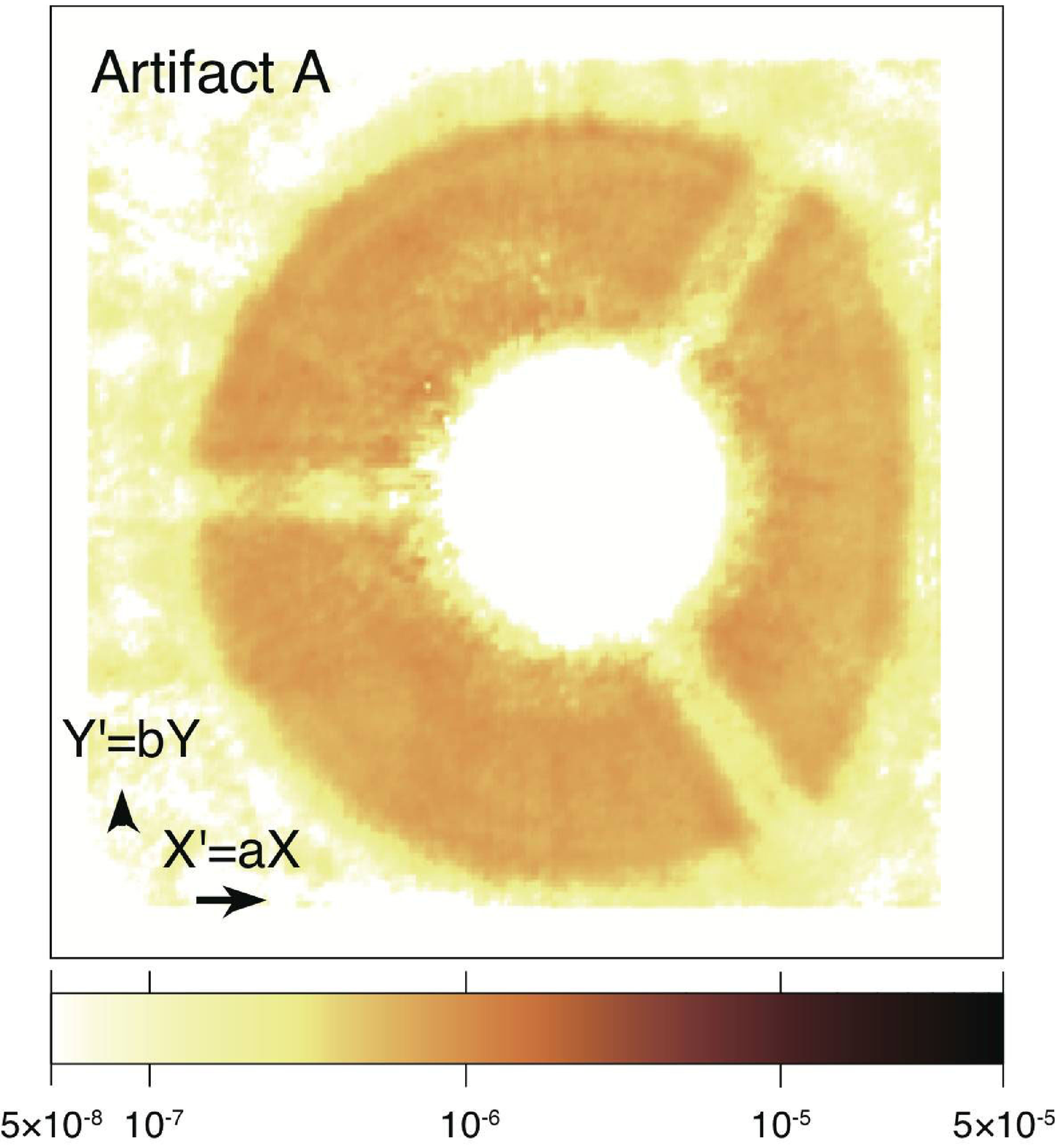}} &
      \resizebox{50mm}{!}{\includegraphics[angle=0]{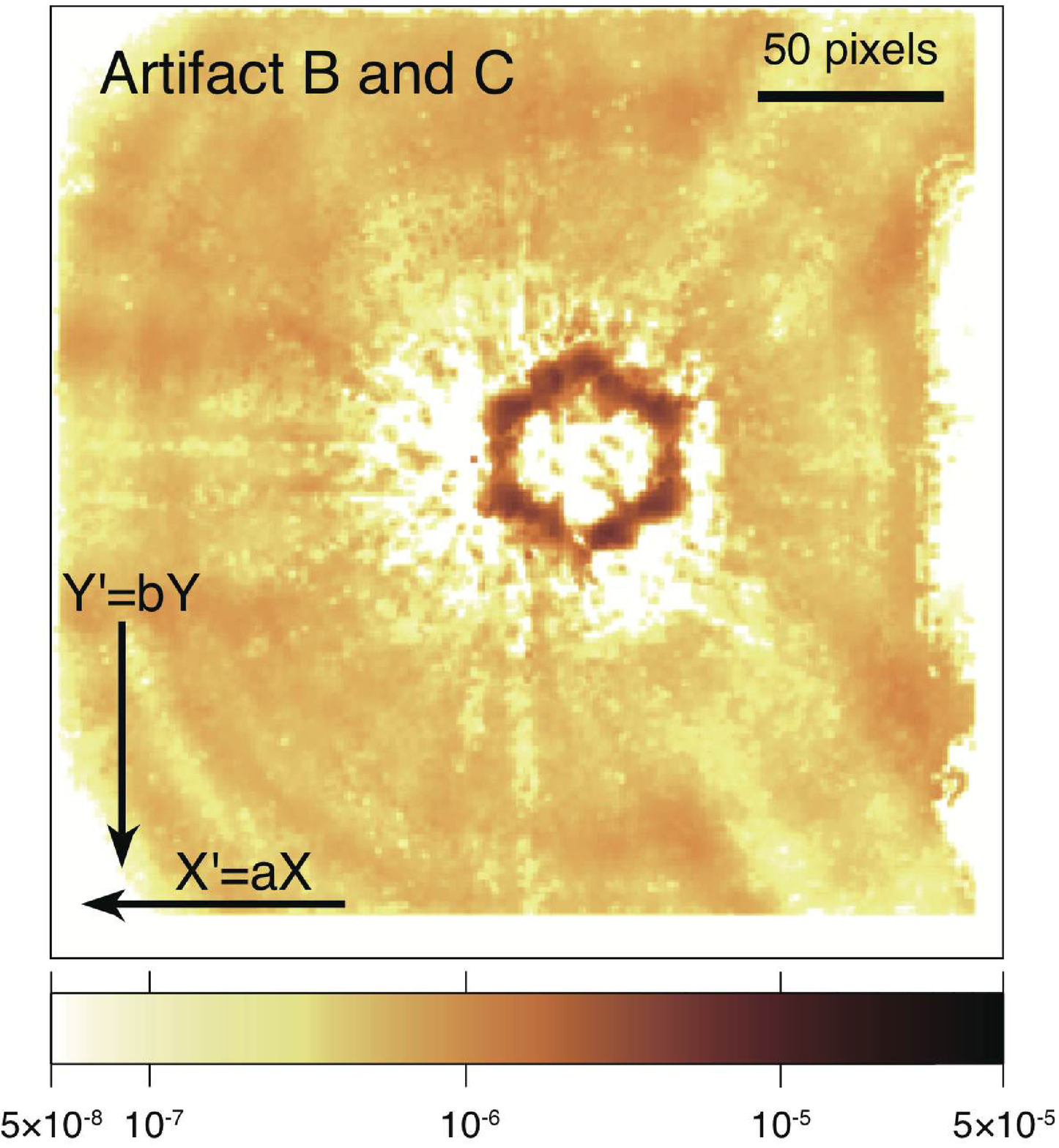}} \\
    \end{tabular}
\caption{The PSF and the artifacts at L15.
The object is located at the center (128 pix, 128 pix).
The arrows in the right two figures indicate the direction of the relative movement. 
All images are displayed in a logarithmic scale with the object signal within 7.5 pixel radius normalized as unity.
\label{figa15}}
%\end{center}
\end{figure}

\begin{figure}[!ht]
%\begin{center}
   \begin{tabular}{ccc}
      \resizebox{50mm}{!}{\includegraphics[angle=0]{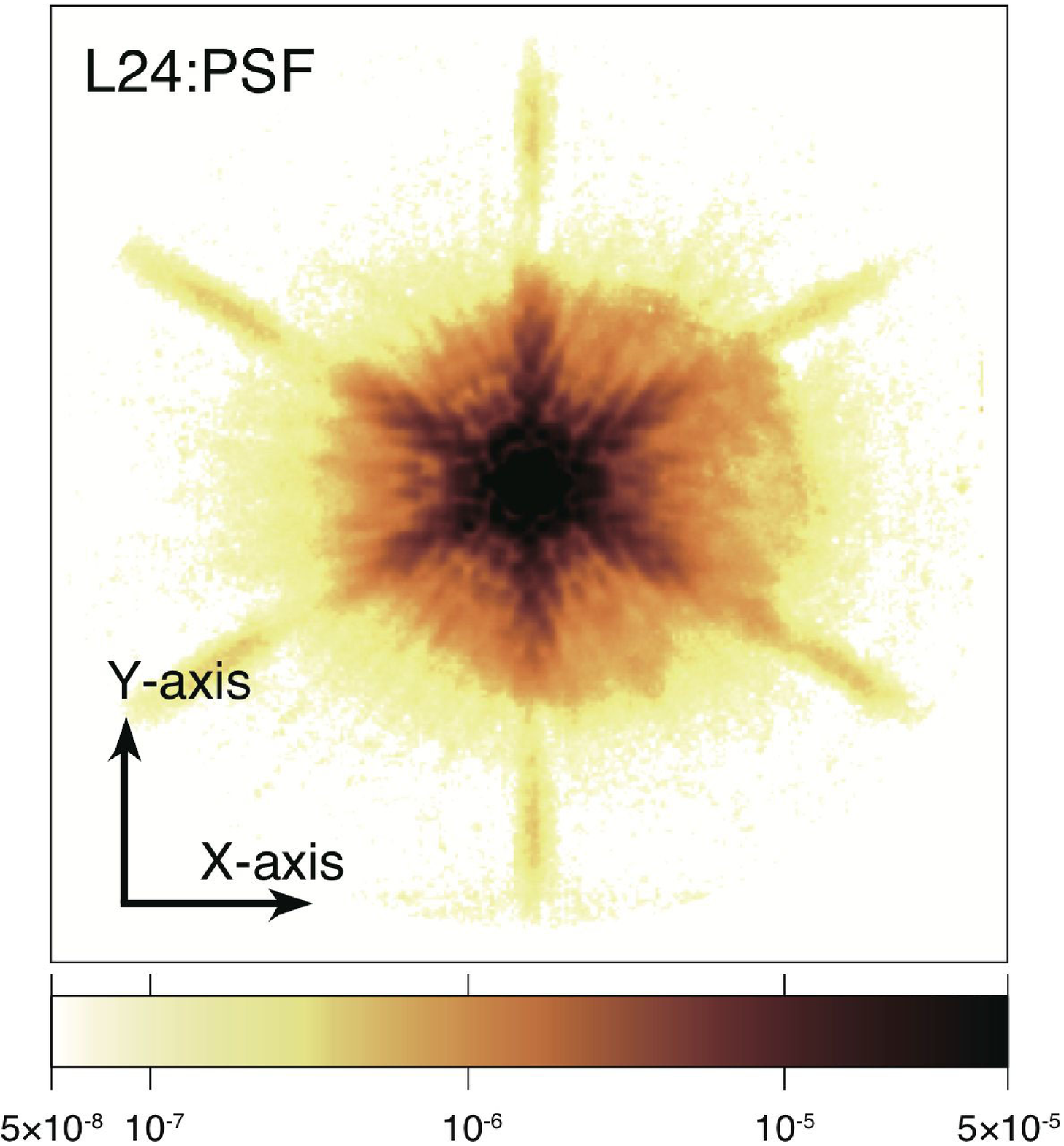}} &
      \resizebox{50mm}{!}{\includegraphics[angle=0]{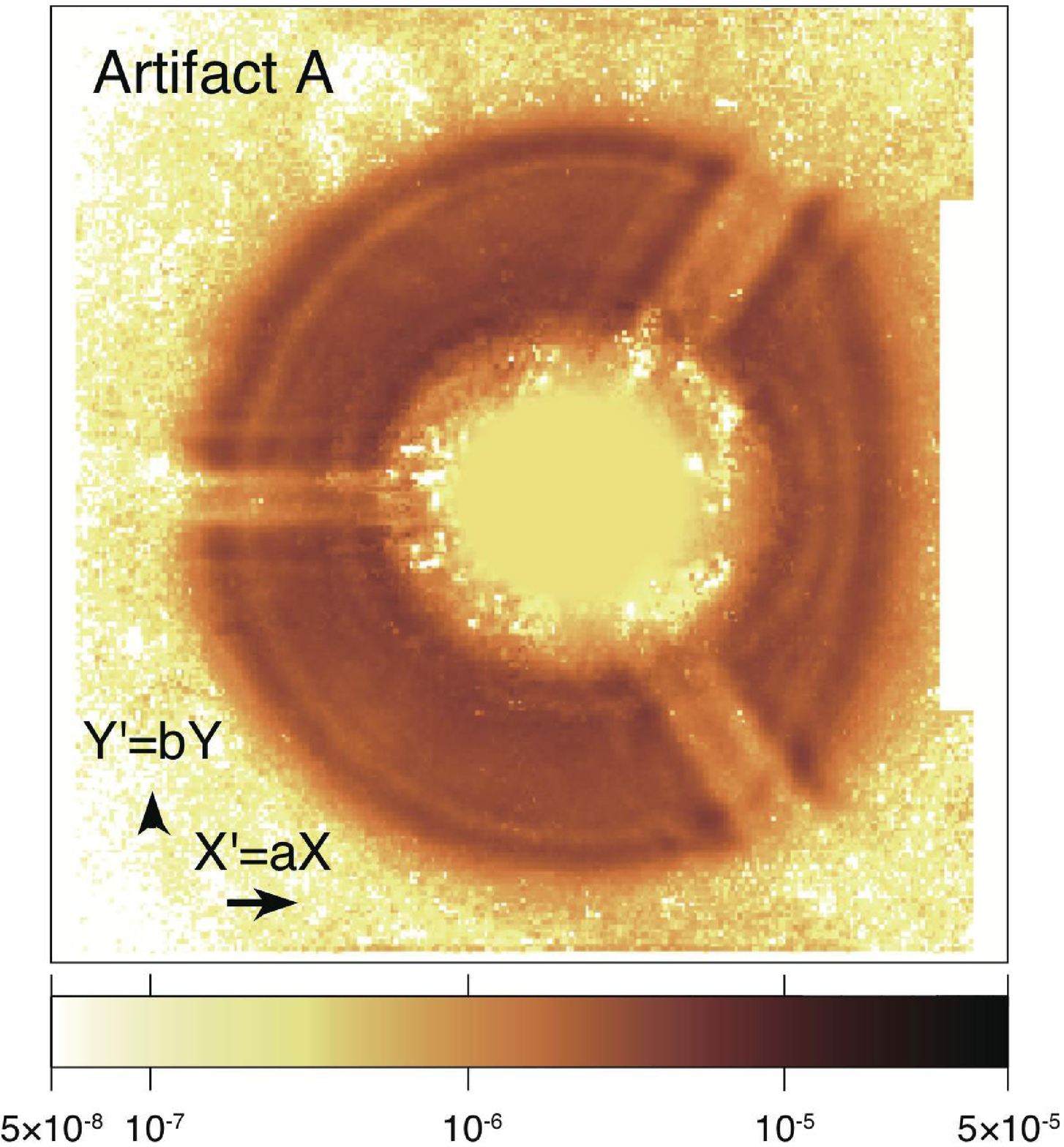}} &
      \resizebox{50mm}{!}{\includegraphics[angle=0]{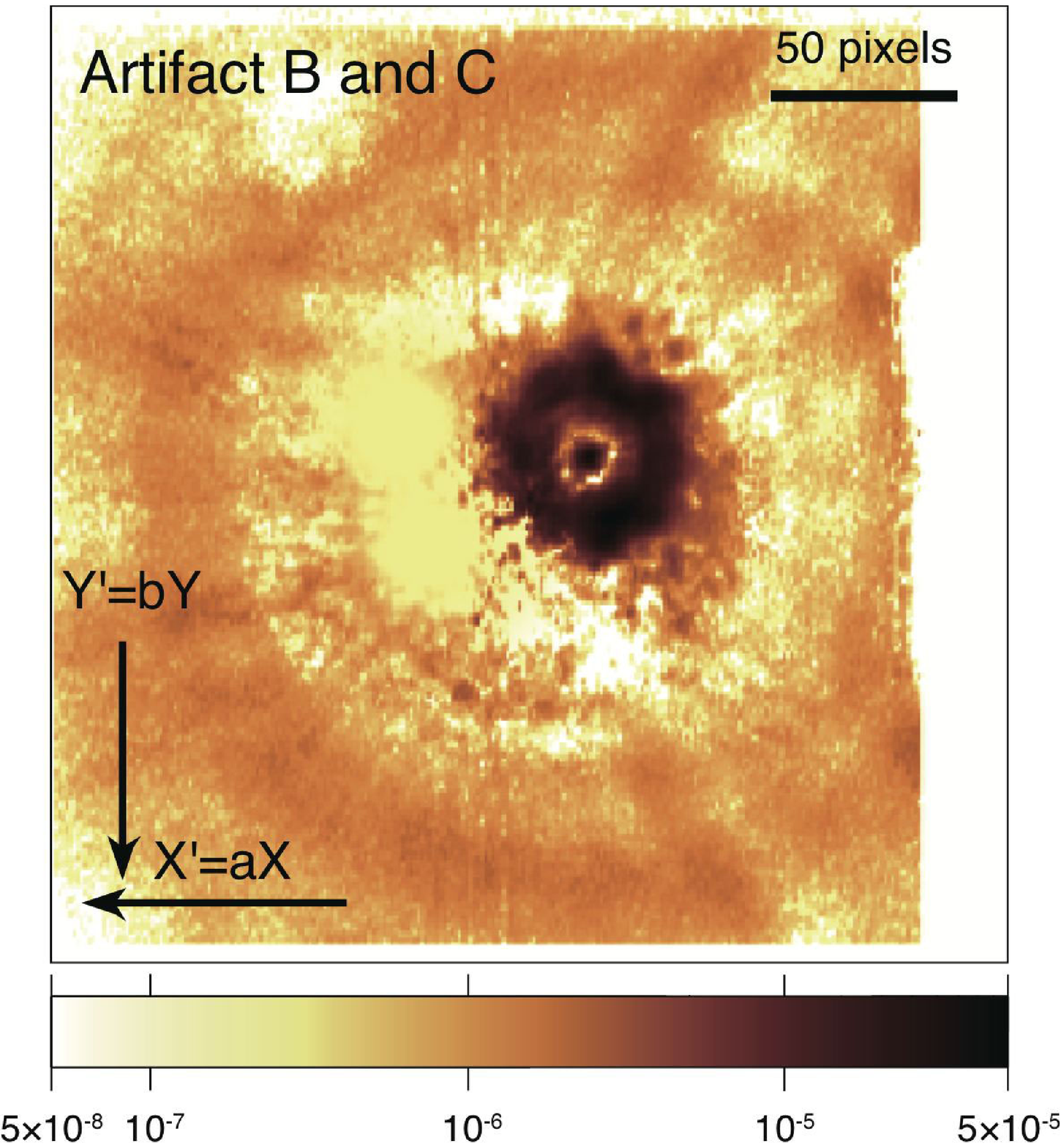}} \\
    \end{tabular}
\caption{The PSF and the artifacts at L24.
The layout is the same as in Figure~\ref{figa15}.
\label{figa24}}
%\end{center}
\end{figure}

The artifacts on L15 images are too faint to derive reliable position movements,
and the fit is made only for L24 images.
We assume the same values of $a$ and $b$ for L15 as derived for L24 since the positions are determined by internal reflections and the spectral dependence is thought to be negligible.
The parameters $a$ and $b$ for artifact A are both positive and thus 
artifact A moves in the same direction as the real object. 
Artifacts B and C move, on the other hand, 
in the opposite direction relative to the real object with the same amount on the image. 
Taking account of  the difference in the movement on the image, we determine the patterns of artifacts A, B, and C, and the PSFs both at L15 and L24
(Figures~\ref{figa15} and \ref{figa24}).
\textcolor{red}{
%Figure 3 shows both the artifacts and the extended PSFs (\S 4).  
The spider patterns and the banding structures from the detectors 
are included in the PSFs (see Figure~\ref{fig.psf}).} 
Because they move with the same amount in the same direction,
artifacts B and C cannot be separated uniquely from the relative movement.
The relative intensities between these two artifacts, however, vary between L15 and L24 
\textcolor{red}{(see Table~\ref{tab:01})}. 
Each artifact image is shifted and stacked by median average, excluding the maximum value
to remove the other artifacts from the output image.
Artifact A in the L15 image is very diffuse and fully overlaps with the extended component of the PSF, 
which makes it difficult to extract.
We assume the same pattern at L24 as for the low frequency ($> 25$ pixels) of artifact A at L15.
Artifact C is much fainter than artifact A at L24
and we assume the same pattern at L24 as at L15.

\begin{figure}[!ht]
\begin{center}
\includegraphics[width=100mm,angle=0]{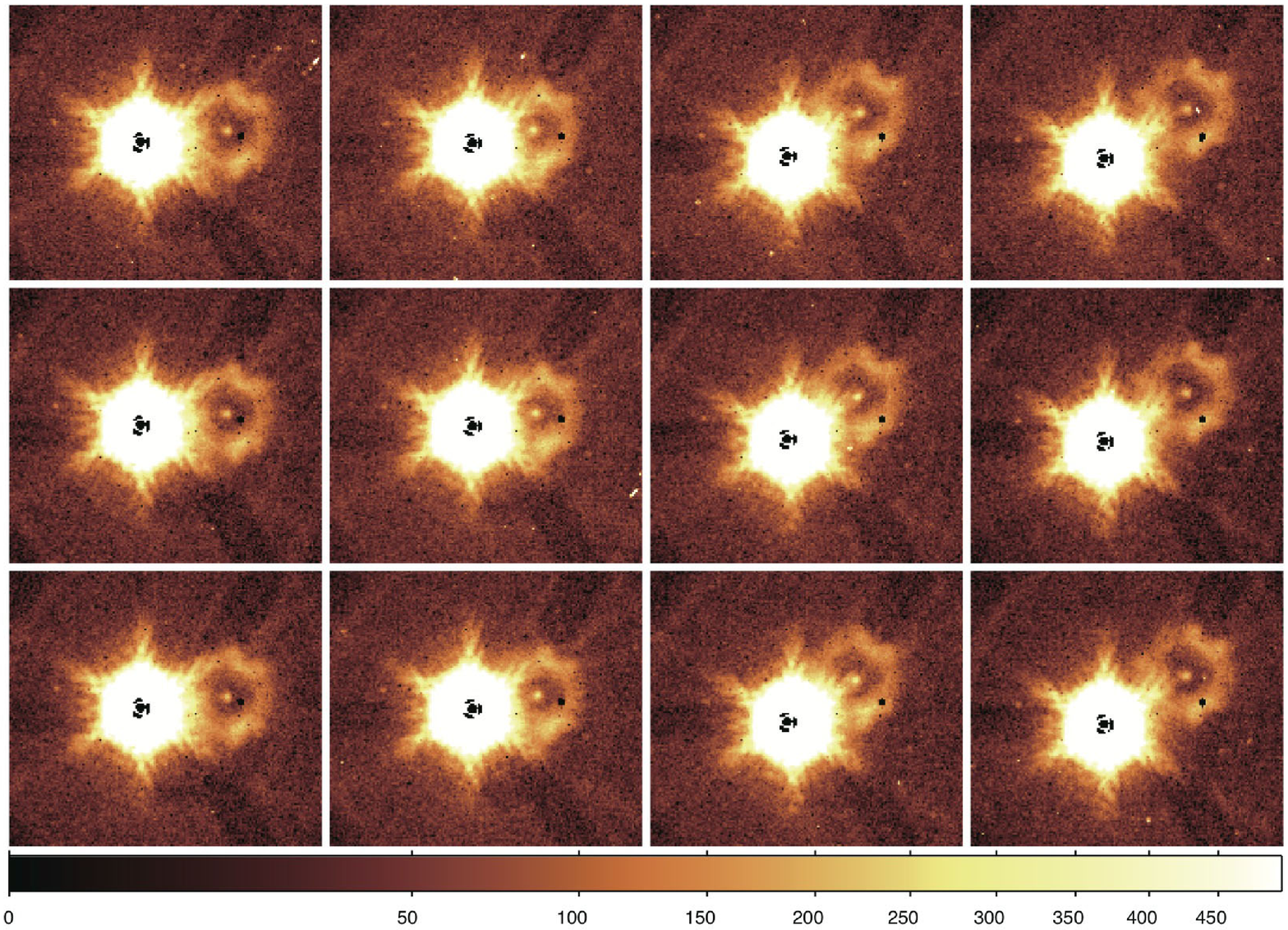}
\caption{Example of the individual dithered images with the artifacts of U Ant at L24 (ID:1710071.1).
The central region of field-of-view (160$\times$150 pixels) is shown.
The color scales are in units of ADU per pixel and all figures are displayed in a logarithmic scale.
\label{fig11}}
\end{center}
%\end{figure}
%\begin{figure}[hhh]
\begin{center}
\includegraphics[width=100mm,angle=0]{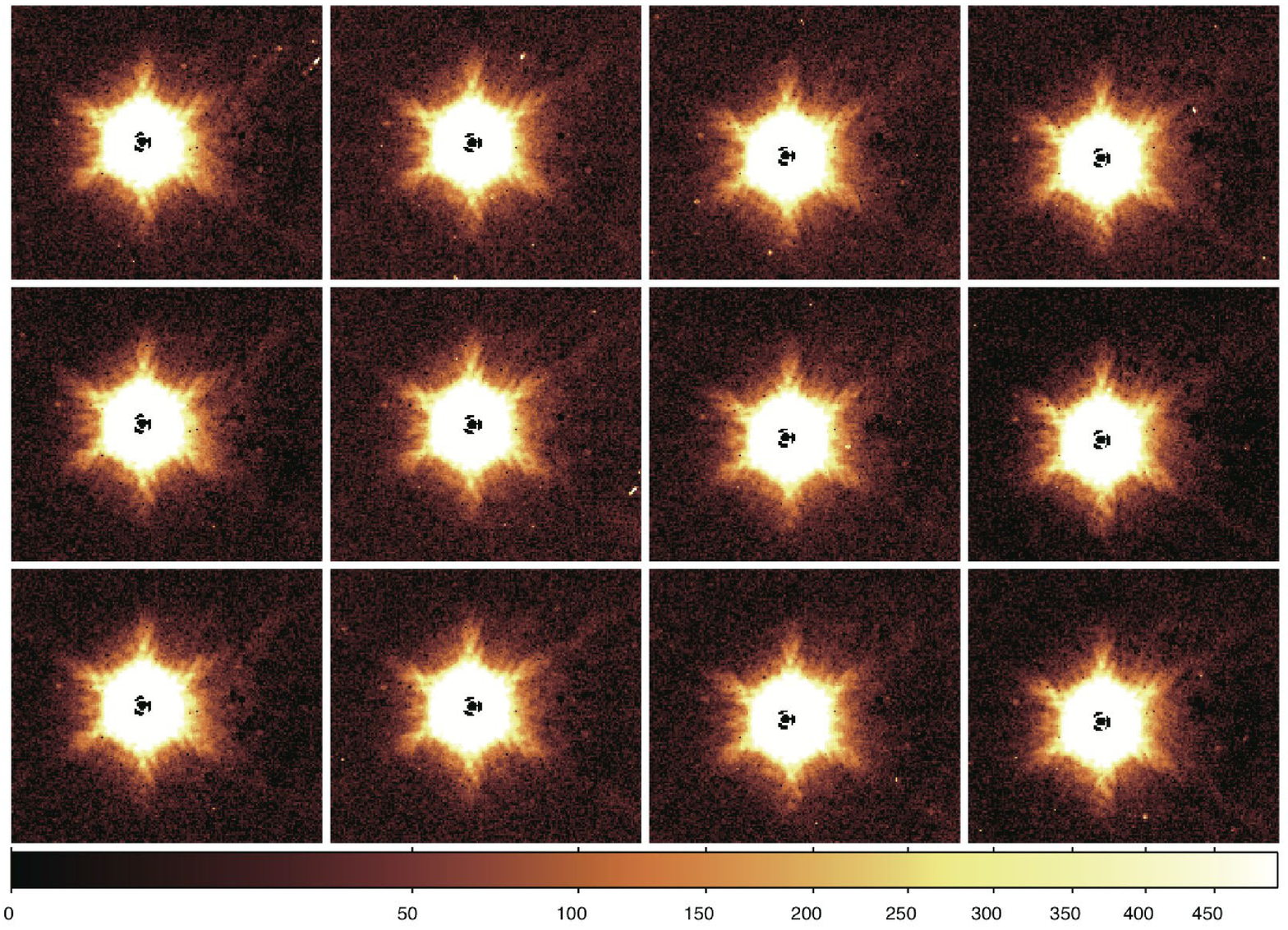}
\caption{Example of the images of U Ant at L24 after the artifacts removal.
All figures are displayed in a logarithmic scale. \label{fig12}}
\end{center}
\end{figure}

The characteristics of artifacts A, B, and C are summarized in Table~\ref{tab:01}.
\textcolor{red}{The total contributions of the artifacts to the input signals are 7.3 \% and 16.1 \% at L15 and L24, respectively.}
The spectral dependence of the intensity of the artifacts relative to the original object cannot be investigated with accuracy because of the lack of sufficient data.
We assume fixed relative intensities of the artifacts as given in Table~\ref{tab:01},
where the artifact signals are normalized by the signal of the target point source within a radius of 7.5 pixels (see below).
Among the MIR-L images we have investigated, the spectral dependence of the relative intensity of the artifacts is less than 15 \%.
Corrections for the artifacts have to be made for each dithered image since the artifacts move differently from real objects.
Figures~\ref{fig11} and \ref{fig12} show examples of the original images with the artifacts and those of the artifacts removed for the carbon star U Antliae at L24, respectively (pointing ID:1710071.1, \citealt{arimatsu11}).
They show that the artifacts seen in Figure~\ref{fig11} are subtracted almost perfectly
 after the removal process (Figure~\ref{fig12}), 
demonstrating the
applicability of
the artifact patterns at L15 and L24 derived in the present study.

\subsection{Extended PSFs of IRC images \label{psf}}

\subsubsection{Extended PSFs \label{sub_psf}}
\textcolor{red}{In general, the PSFs of mid-infrared instruments tend to be extended due to the diffraction and the scattering within the detector arrays \citep{hola04}.}
We determine the PSFs over the entire field-of-view of the four imaging bands (S7, S11, L15, L24)
by subtracting the artifacts
from calibration and science observations of stars.
The observations used to determine the PSFs are listed in Table~\ref{tab4}.
To obtain HDR PSFs, we use targets  with a wide range of brightness. %of various brightnesses.
We also employ short-exposure and long-exposure images,
the latter of which was taken with 28 times longer exposure time than the former.
%We compared the 
Images after the artifact subtraction at each dithering position are shifted and combined with an accuracy of 0.1 pixels 
to produce a brightness map, and we determine the extended component of the PSFs down to $10^{-6}$ of the total intensity.
Figure~\ref{fig.psf} shows the PSFs, which indicates significant morphological differences among the imaging bands.
Especially in the PSFs at S7 and L15, a banding structure along the cross-scan direction (hereafter, defined as the X-direction) is seen 
in addition to the structures produced by the secondary mirror spiders separated by 120 degrees.
A similar banding pattern is seen in {\it Spitzer}/IRAC images, which is attributed to light scattering inside the detector \citep{pipher04}.
The banding shows a strong spectral dependence and the morphological differences can be attributed partly to the internal reflections in the Si:As detector 
and/or to diffraction of light.
\textcolor{red}{Uniform offsets that appear under high incoming photon conditions \citep{pipher04} are not recognized in the IRC PSF object images, when they are compared with the reference sky images (\S~\ref{sub_Lartifact}).}

The full width of half maximums (FWHMs) of the obtained PSFs are listed in Table~\ref{fwhm}.
The obtained values of the FWHMs are smaller than the published values (Table 2 of \citealt{onaka07}) at the short wavelength channels.
The present values are consistent with the diffraction-limited performance at 7.3\,$\mu$m \citep{kaneda07}.
The published values are derived from a number of stellar images and the averaging process could increase the FWHM.
In another word, they provide "average" FWHMs when the standard reduction toolkit is used.
On the other hand, the PSFs in the present study are derived from the data that are carefully shifted and stacked. 
This process requires for the data with sufficient signal-to-noise ratios.
When the present HDR PSF is applied to a standard processed image,
the inner part of the new PSFs needs to be adjusted to the individual image,
which might be degraded by a less accurate stacking process, 
while the outer part provides proper information on the extended component of the PSFs.
\subsubsection{Aperture Corrections  \label{sub_AC}}

The extended component of the PSFs can be characterized quantitatively by the encircled energy.
It is also useful for practical applications to calculate the aperture correction factor $AF$, 
which is defined as the reciprocal of the encircled energy normalized at a given radius.
We set the reference radius as 7.5 pixels,
with which the absolute calibration of the MIR-S and MIR-L data is carried out \citep{tanabe08}.
This radius corresponds to 17\arcsec.55 for the MIR-S and 17\arcsec.88 for the MIR-L channel, respectively.
According to \citet{Jarrett06}, we approximate $AF$ as a function of the radius $r$ from the center of the object as 
\begin{eqnarray}
AF (r) = A  \exp (-r^B)  + C,
\label{eqAC}
\end{eqnarray}
where $r$ is in pixels, 
and $A$, $B$ and $C$ are the fitting parameters.
They are given in Table~\ref{tab:a04} and $AF(r)$ is plotted in Figure~\ref{fig.AF}.
Figure~\ref{fig.AF} indicates that equation~(\ref{eqAC}) provides good fits for $r > 7.5$ pixels.
%Photometry of extended sources is subject to a diffuse correction factor because of the extended PSFs.
%The diffuse calibration factor is defined as following equation:
%\begin{eqnarray}
%{\rm Surface\ Brightness} = {\rm Observed\ Surface\ Brightness} \times {\rm Diffuse\ Calibration\ Factor}
%\end{eqnarray}
%Assuming a diffuse emission that extends flattery over the entire field-of-view, 
%the observed flux density corresponds to be the flux density of point source which integrated over the entire field-of view.
%Thus the correction factor for the infinite extended source should be the aperture correction coefficient $C$.
%The obtained diffuse calibration factors for the IRC wavebands are listed in Table.~\ref{tab:a06}.

The calibration for diffuse emission needs to take account of the fact that a large fraction of the incident photons 
spread over the detector array.
The correction for diffuse emission can be given by equation~(\ref{eqAC}) with $r \to \infty$.
Thus parameter $C$ gives the correction factor for diffuse emission. 
As indicated in Table~\ref{tab:a04}, the correction factor has spectral dependence.
Scattering within the detector array is known to have strong spectral dependence, 
peaking in the 5--6\,$\mu$m region \citep{pipher04}.
It explains part of the dependence in the MIR-S bands.
For the MIR-L bands, diffraction of light is likely to make a significant contribution.

\subsection{Revised Flat Frames for MIR-L images \label{flat}}
\begin{figure}[!h]
    \begin{tabular}{cc}
      \resizebox{70mm}{!}{\includegraphics[angle=0]{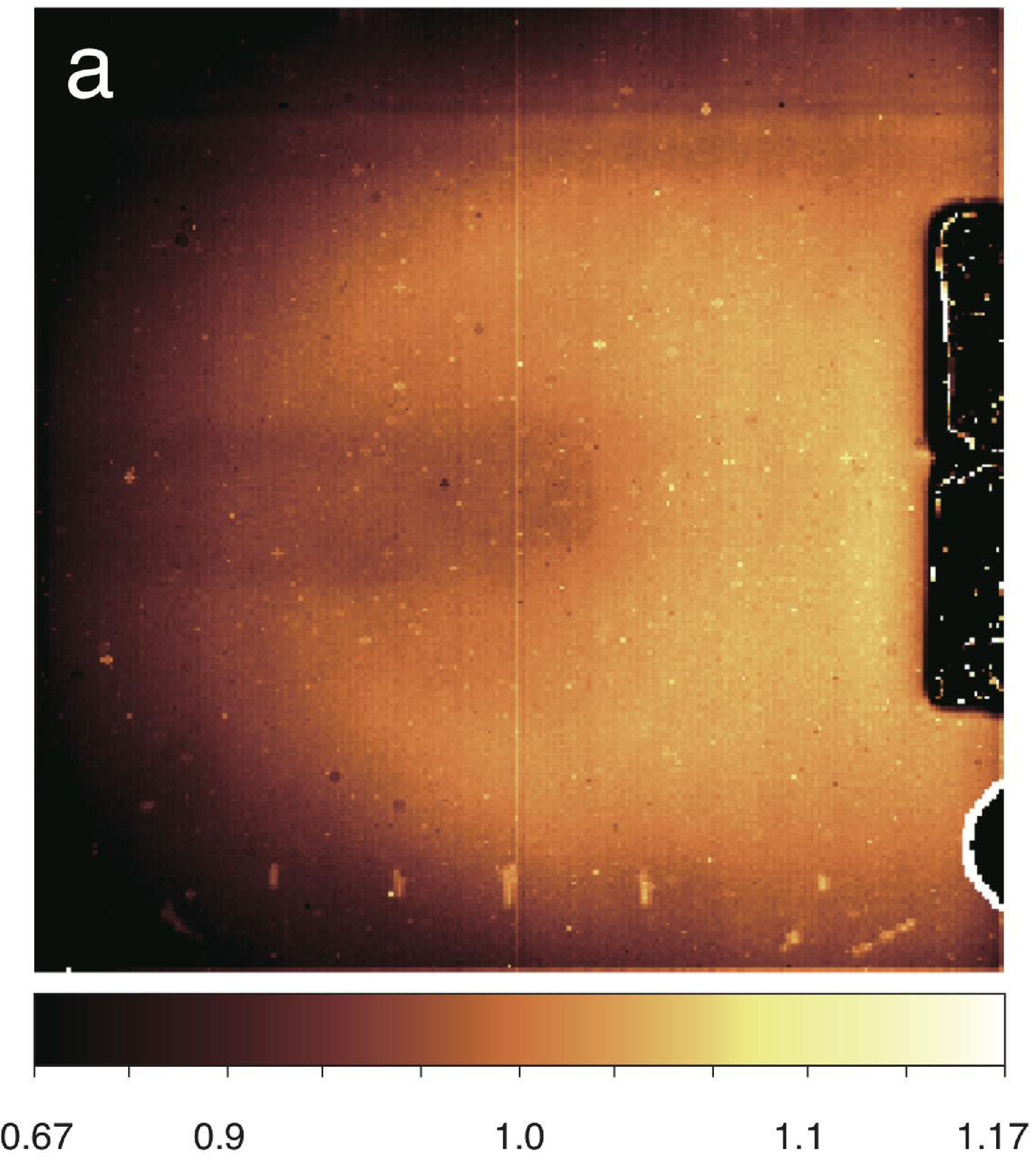}} &
      \resizebox{70mm}{!}{\includegraphics[angle=0]{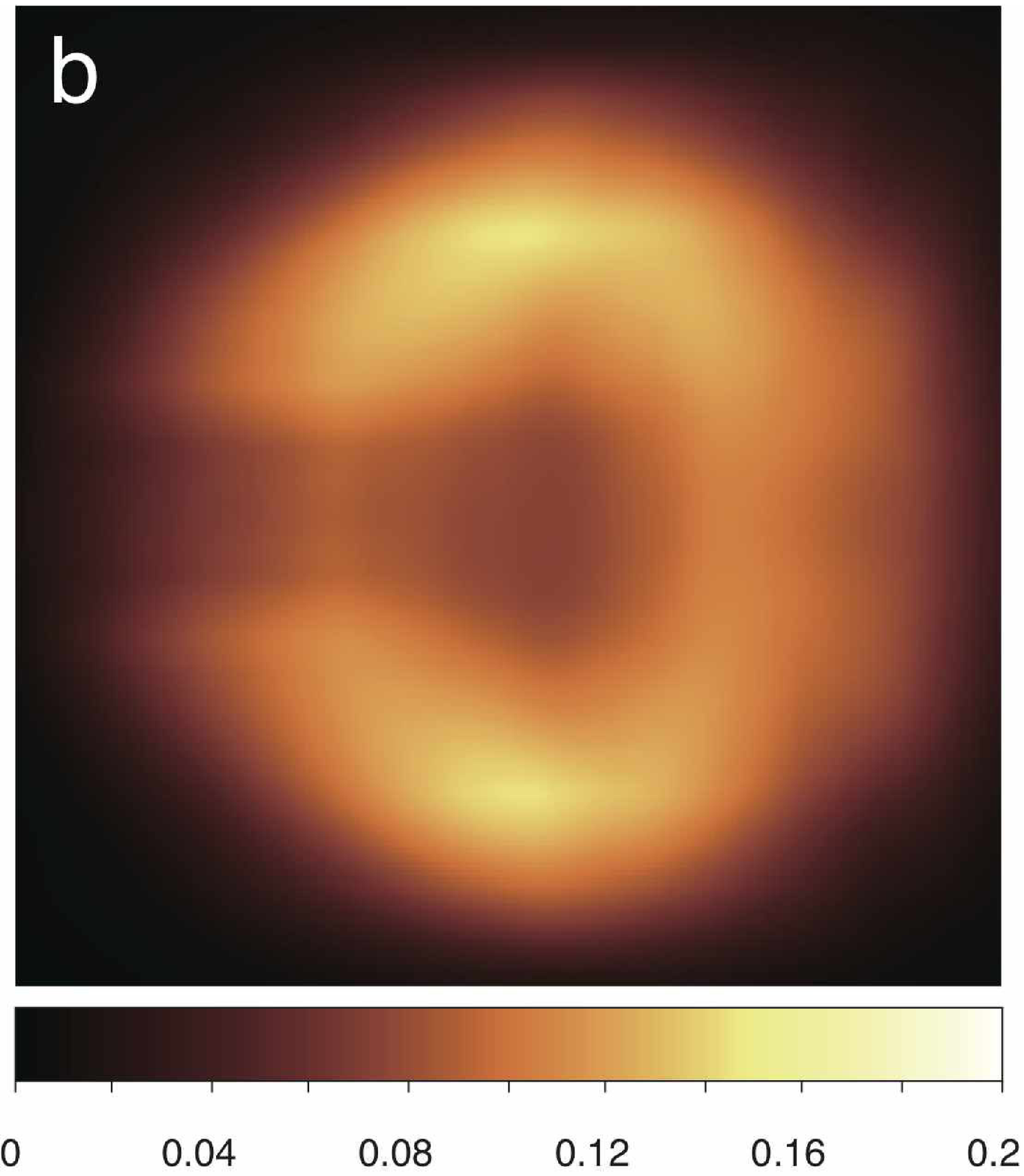}} \\
      \resizebox{70mm}{!}{\includegraphics[angle=0]{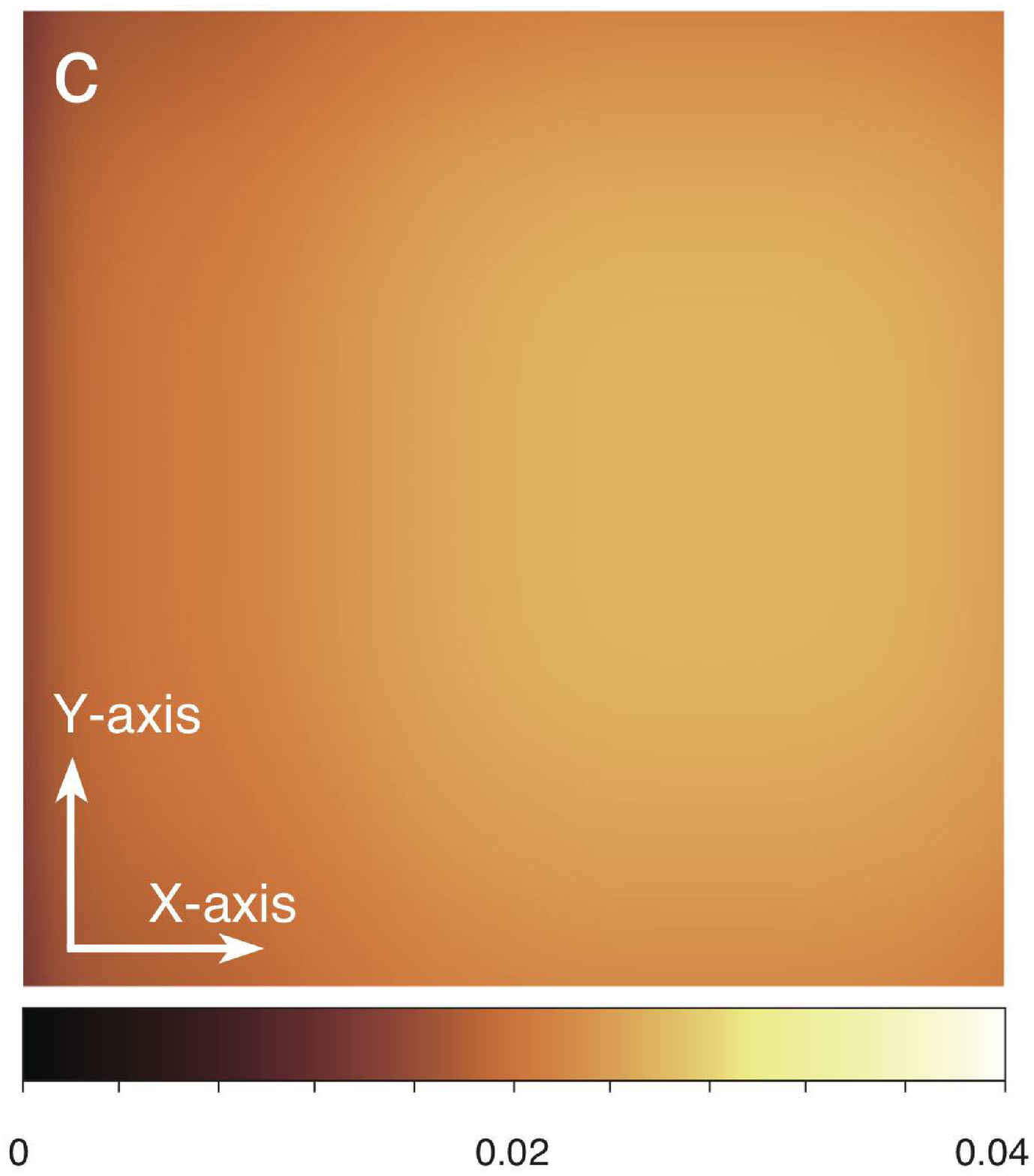}} &
      \resizebox{70mm}{!}{\includegraphics[angle=0]{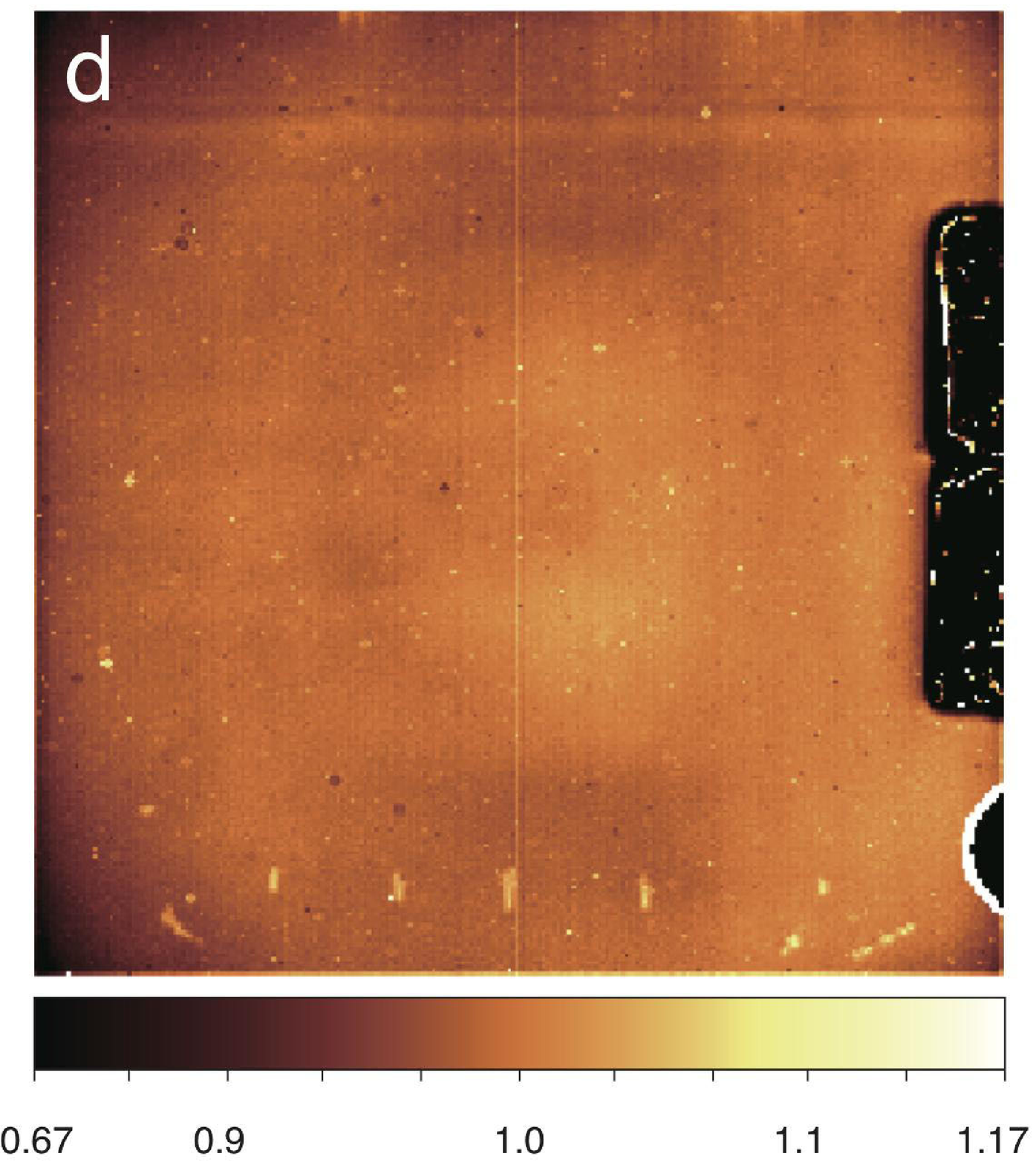}} \\
    \end{tabular}
    \caption{(a) Previous flat data of L24,
(b) artifact A predicted from a uniform background,
(c) artifact B and C patterns predicted from a uniform sky,
and (d) flat data of L24 corrected for the artifacts.
The entire detector array (256 $\times$ 256 pixels) is shown.}
    \label{fig_flat_im}
\end{figure}

Diffuse emission images are affected not only by the artifacts 
\textcolor{red}{and the extended PSF components} 
produced by bright objects in the field-of-view,
but also by those produced by diffuse sky background.
Signals of the artifacts produced by the diffuse background in each pixel are small,
but the total amount of the signals integrated over the entire array becomes significant.
\textcolor{red}{Flat data for mid-infrared instruments are usually produced from the sky background data.}
Previous flat data for the MIR-L channel are created from a large number of background sky images.
\textcolor{red}{Thus the artifacts at L15 and L24 can affect the derived flat data}.
In fact, they show strange patterns, which are attributed to the artifact effects \citep{lorente08}.
The flat data show a steep decline at the edges of the field-of-view (Figure~\ref{fig_flat_im}a).
It is most likely that the decline is produced by the effect of the artifacts, especially artifact A. 

\begin{figure}[!h]
\begin{center}
\includegraphics[width=\hsize,angle=0]{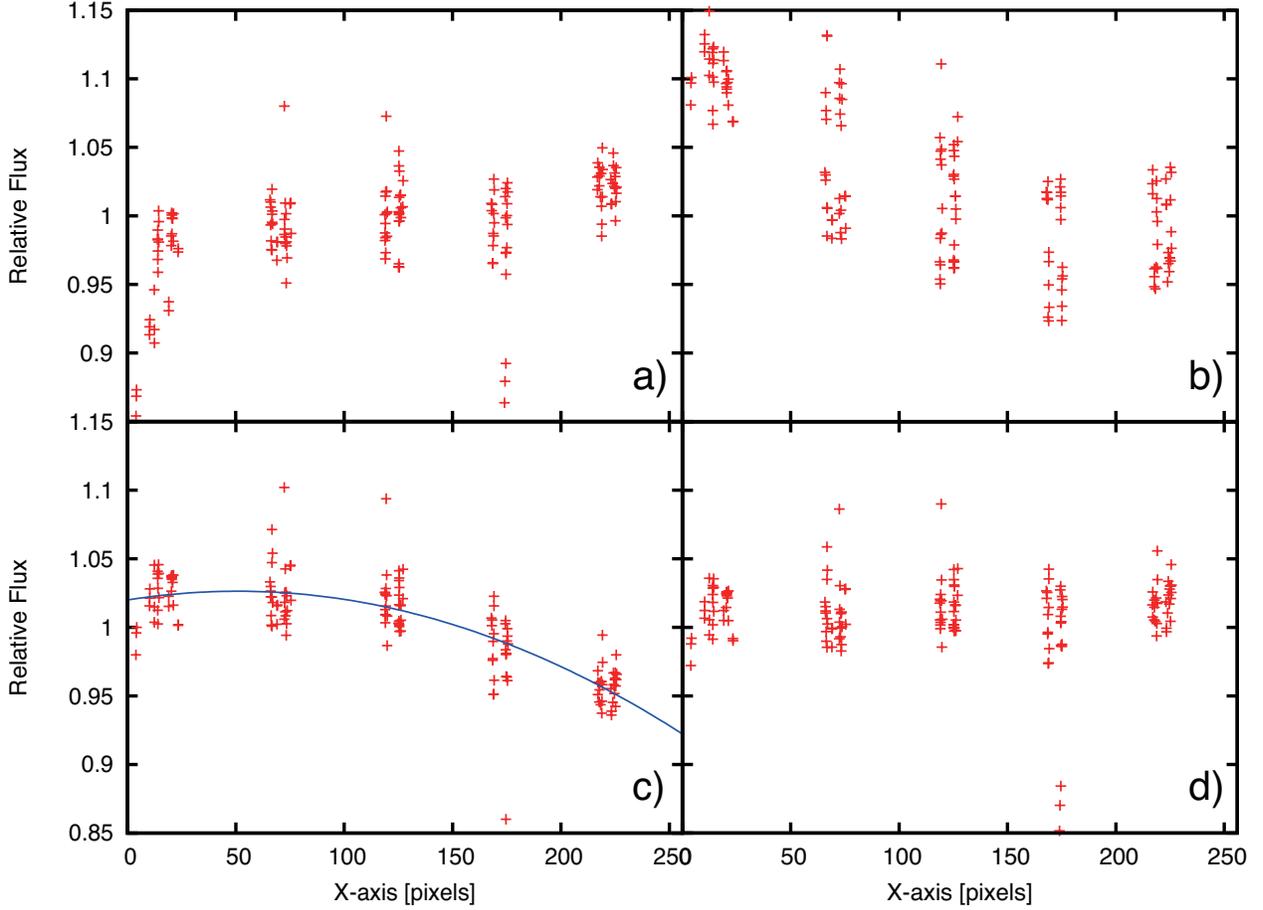}
\end{center}
%    \begin{tabular}{cc}
%      \resizebox{75mm}{!}{\includegraphics[angle=0]{flat0.eps}} &
%      \resizebox{75mm}{!}{\includegraphics[angle=0]{flat01.eps}} \\
%      \resizebox{75mm}{!}{\includegraphics[angle=0]{flat_0_1.eps}} &
%      \resizebox{75mm}{!}{\includegraphics[angle=0]{flat02.eps}}\\
%    \end{tabular}
    \caption{Comparison of the aperture photometry of the same point source (F06009$-$6636) at 25 different positions at the L24 band.
All the plots are made along the X-axis of the image.
(a) The photometric data without flat-fielding.
(b) Those corrected with the previous flat data.
(c) Those corrected with the artifact-subtracted flat data. 
(d) Those corrected with the artifact-subtracted flat data with the slope correction shown in Figure c (see text). 
The photometry is all carried out with the standard aperture size (7.5 pixels in radius)
and the flux is normalized by the average of the fluxes of the data taken at the center position.}
    \label{fig_flat_x}
\end{figure}

\begin{figure}[!h]
\begin{center}
\includegraphics[width=\hsize,angle=0]{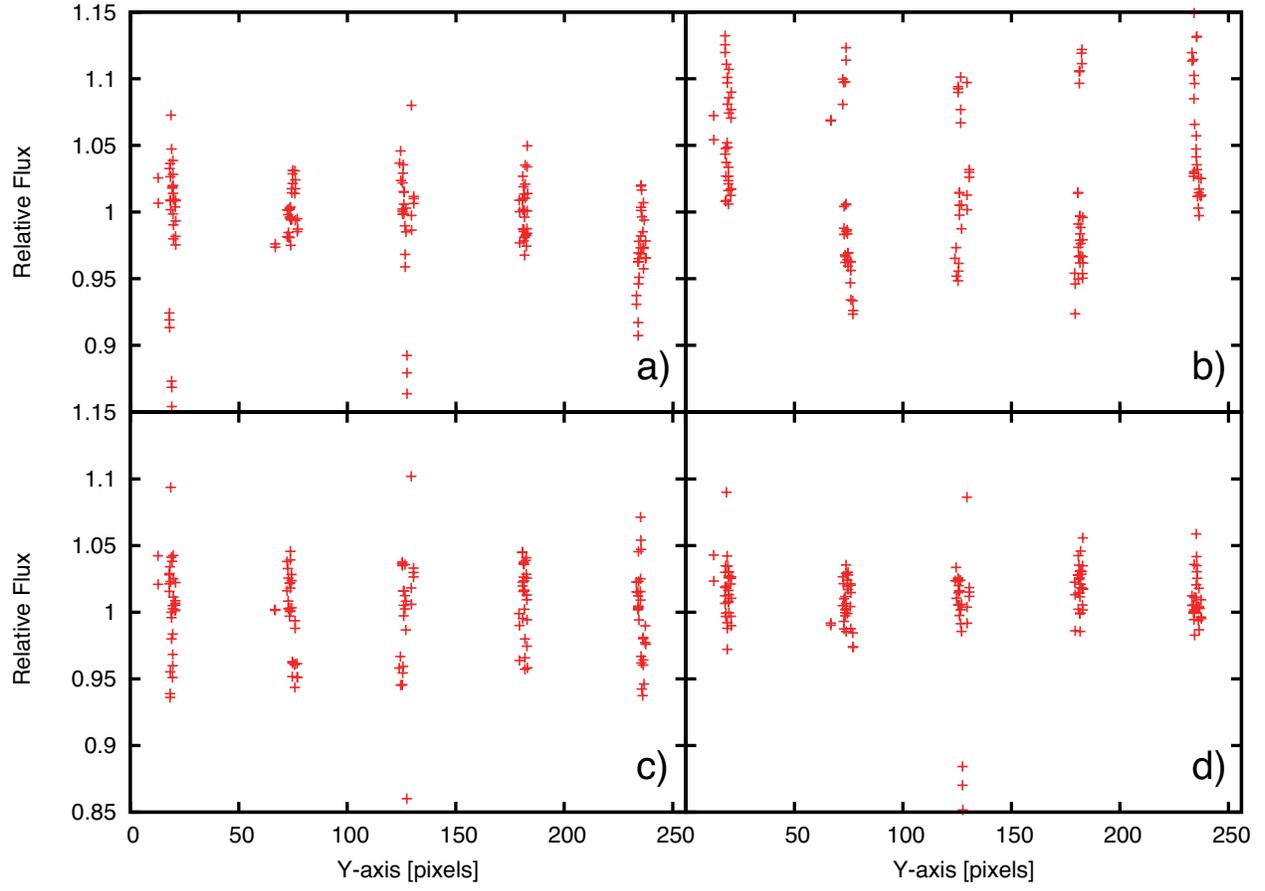}
\end{center}
%    \begin{tabular}{cc}
%      \resizebox{75mm}{!}{\includegraphics[angle=0]{flat0y.eps}} &
%      \resizebox{75mm}{!}{\includegraphics[angle=0]{flat01y.eps}} \\
%      \resizebox{75mm}{!}{\includegraphics[angle=0]{flat0_1y.eps}} &
%      \resizebox{75mm}{!}{\includegraphics[angle=0]{flat02y.eps}}\\
%    \end{tabular}
    \caption{Same as Figure~\ref{fig_flat_x} except that the plots are made along the Y-axis of the image.}
    \label{fig_flat_y}
\end{figure}

\begin{figure}[!ht]
  \begin{tabular}{cc}
      \resizebox{80mm}{!}{\includegraphics[angle=0]{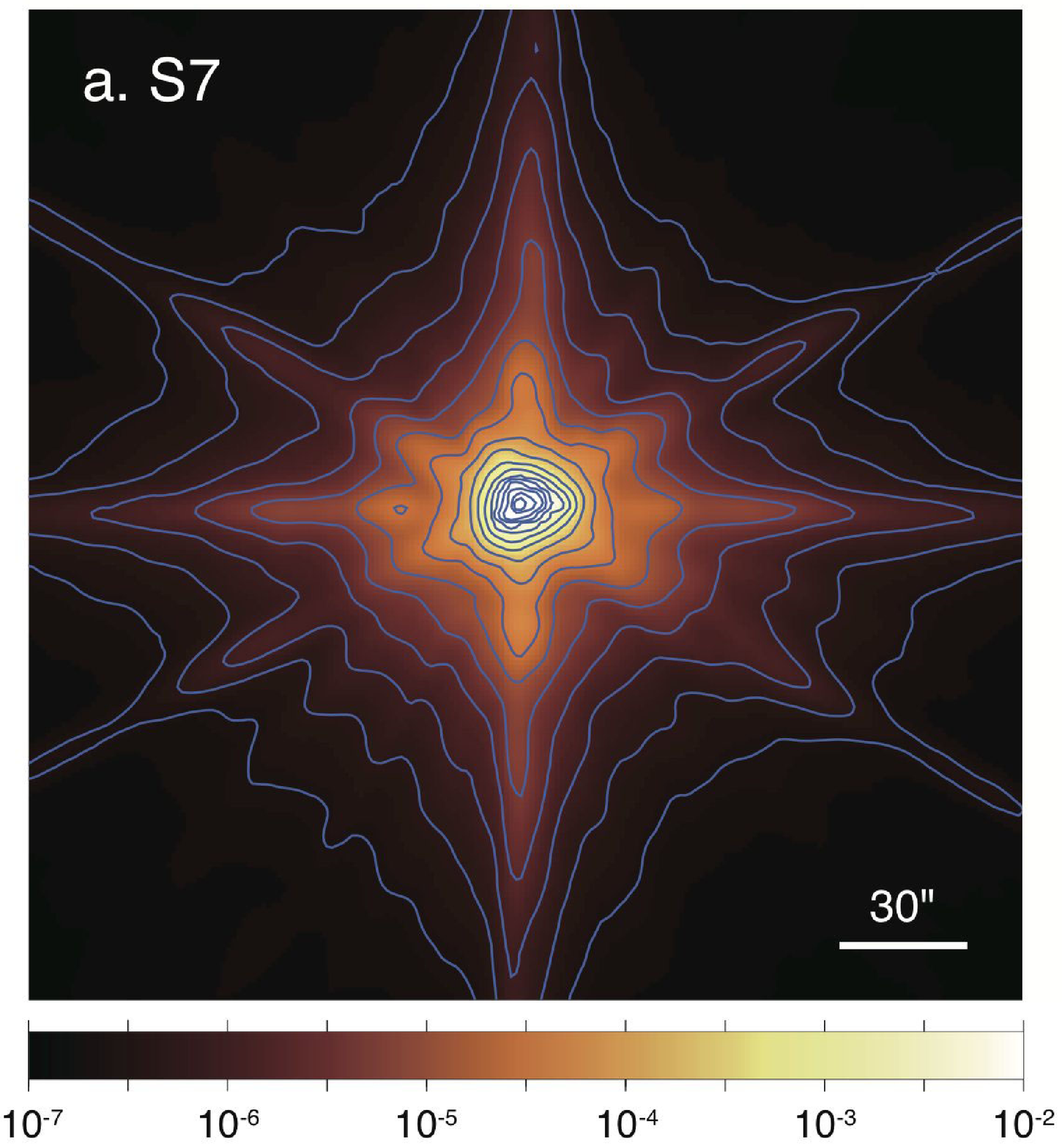}} &
      \resizebox{80mm}{!}{\includegraphics[angle=0]{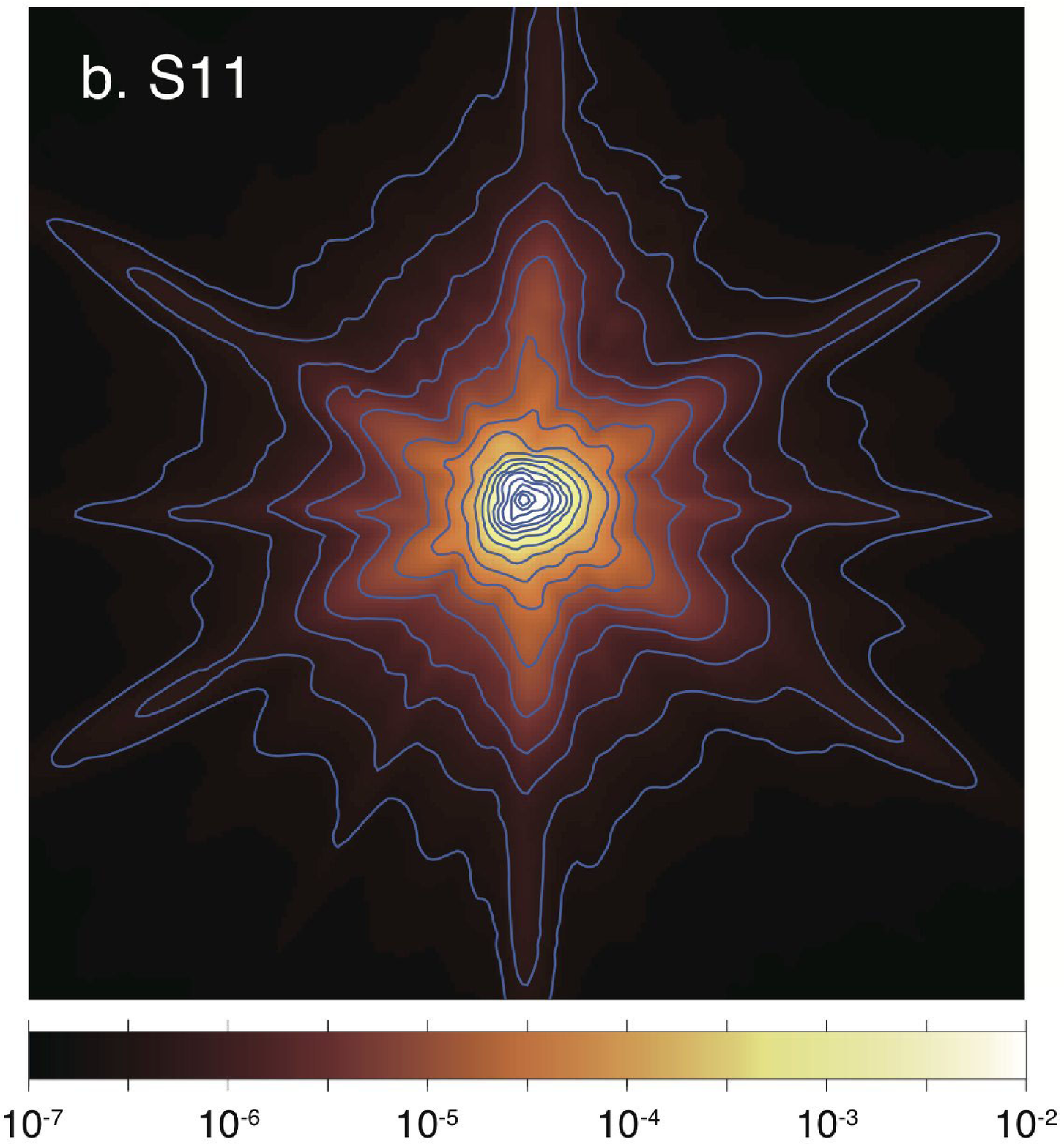}} \\
      \resizebox{80mm}{!}{\includegraphics[angle=0]{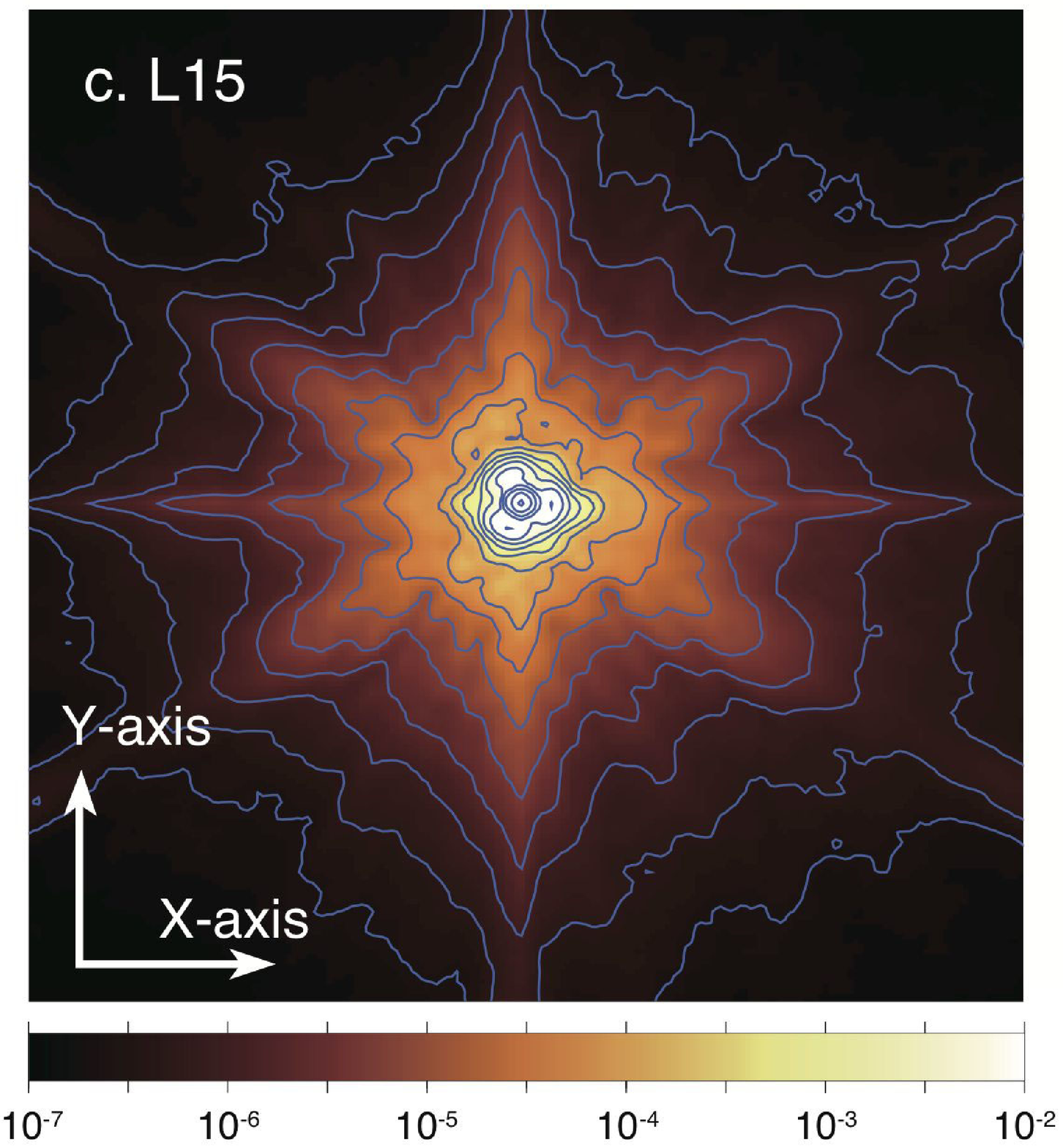}} &
      \resizebox{80mm}{!}{\includegraphics[angle=0]{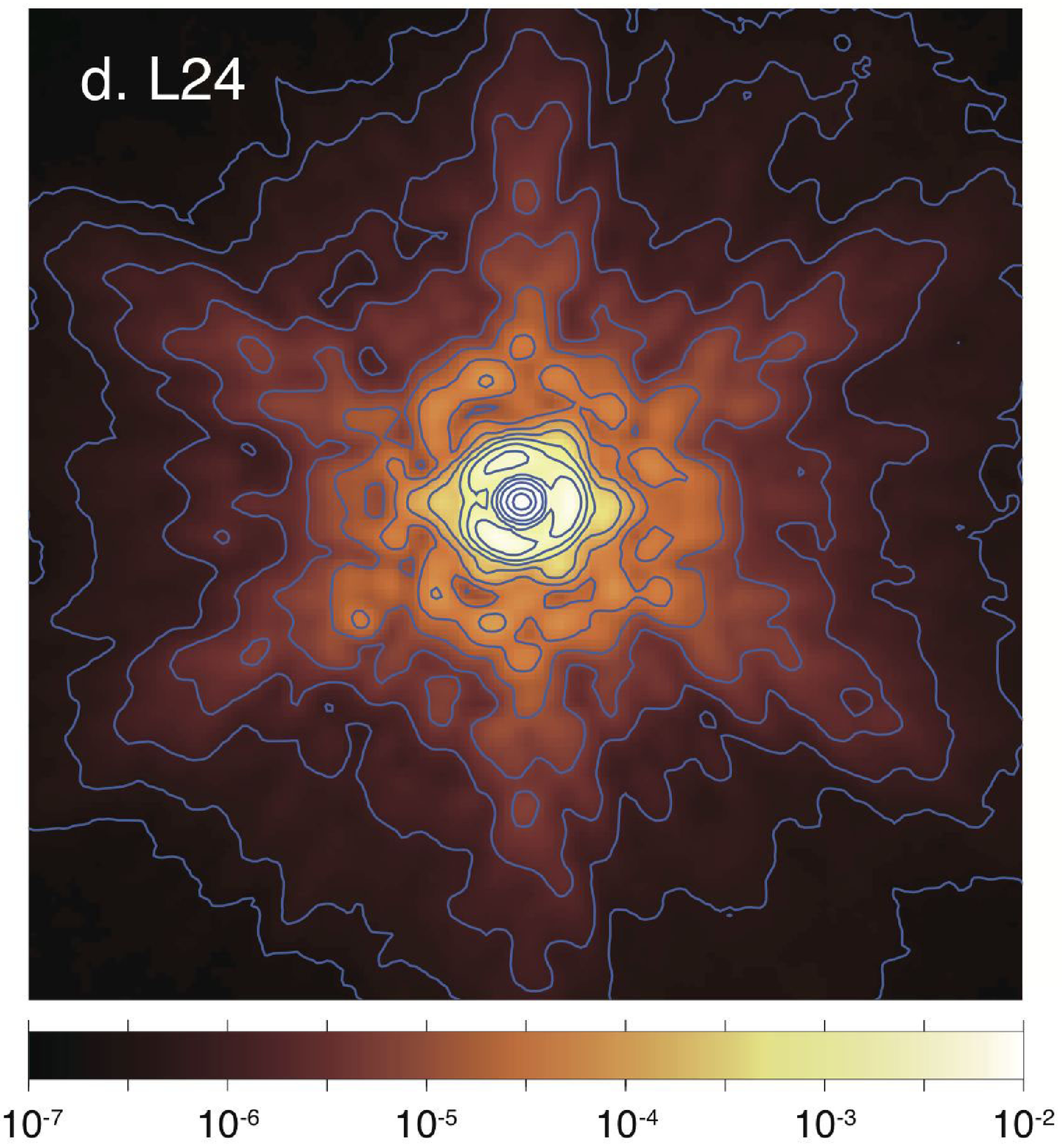}} \\
    \end{tabular}
\caption{PSFs of {\it AKARI}/IRC imaging bands at (a) S7, (b) S11, (c) L15, and (d) L24.
All figures are displayed in a logarithmic scale with the total intensity of unity. 
The contour levels are set as $2^{(n+1)} \times10^{-5}$ of the total intensity.}
\label{fig.psf}
\end{figure}

\begin{figure}[!ht]
\begin{center}
\includegraphics[width=\hsize,angle=0]{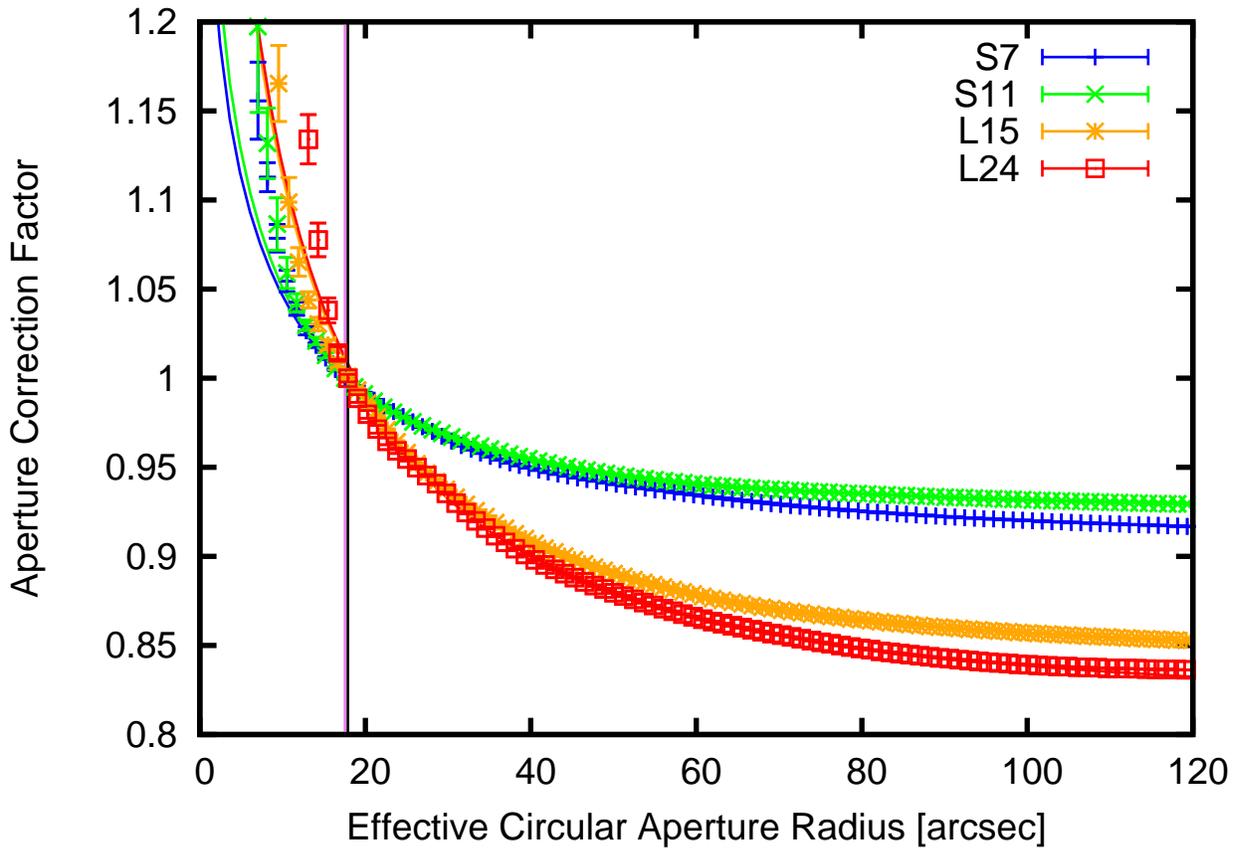}
\caption{Aperture correction factors for the S7 (blue), S11 (green), L15 (orange), and L24 (red) bands. 
The lines are the fitting results with equation (\ref{eqAC}).
The standard photometry radius (7.5 pixels) is shown by the violet and black lines for MIR-S and MIR-L, respectively. \label{fig.AF}}
\end{center}
\end{figure}

To investigate the effects of the artifacts on the flat data, we carried out observations of the MIR-L
standard star IRAS F06009$-$6636 at 25 different positions of a 5$\times$5 grid separated by $\sim 2\arcmin$
in the field-of-view of the MIR-L (Table~\ref{tab4}) and compare the variation in the signal with the flat data.
The flux of the star is measured with aperture photometry with a radius of 7\arcsec.5 and is derived for each dithered image.
The comparison indicates that the previous flat data overcorrect the flux of a point source at an edge of the field-of-view by 15 \% at most at L24 
\textcolor{red}{(Figures~\ref{fig_flat_x}b and \ref{fig_flat_y}b)}.
The standard deviation of the flux variation increases from 3.8 \% (without flat-fielding) to 6.4 \% if the flat correction is made.
These observations were useful to estimate the accuracy of the previous flat-fielding process of MIR-L images,
but did not have sufficient spatial information to produce new flat data.

We produce new flat data for L15 and L24 from the previous flat data by correcting for the artifacts derived in \S~\ref{sub_Lartifact}.
First we assume that the sky background is uniform over the entire field-of-view 
and integrate the contribution from the signal of each position to obtain the artifact pattern that is produced by the uniform background.
Next the artifact pattern is searched for in the previous flat data and the amplitude of the artifact is estimated.
The scaled artifact pattern is then subtracted from the previous flat data.
The  artifact-subtracted-flat data are renormalized at the center position to keep the present absolute calibration valid 
since most of the standard stars are observed around the center of the field-of-view \citep{tanabe08}.
Finally we apply the newly-derived flat data to the 5$\times$5 standard star observations.
The results indicate that the flat is significantly improved in the Y-direction 
\textcolor{red}{(Figure~\ref{fig_flat_y}c)}, 
but there still remains a shallow decline around the edge in the X-direction 
\textcolor{red}{(Figure~\ref{fig_flat_x}c)}.
There may still be a very diffuse artifact that remains uncorrected.
We fit the pattern in the X-direction with a quadratic function of the position 
and apply it to the flat data to create the final flat data (Figure~\ref{fig_flat_im}d).
With the final artifact-subtracted and slope-corrected flat data, the standard deviation of the flux variation of the standard star decreases from 3.8 \% to 2.8 \% for L24 
\textcolor{red}{(see Figures~\ref{fig_flat_x}d and \ref{fig_flat_y}d)}.

For the L15 band, a similar trend is seen in the previous flat data.
The same procedure is applied as for the L24 band to obtain new flat data after the slope correction in the X-direction.
The standard deviation of the flux variation of the standard star
improves from 3.2 \% (without flat-fielding) to 1.2 \% (with the
artifacts-subtracted flat data with the slope correction) for the L15 band.

%The new flat images for L15 and L24 are included in the latest version of the IRC imaging reduction toolkit, 
%which is available via the {\it AKARI} observers page (http://www.ir.isas.jaxa.jp/ASTRO-F/Observation). 
%The validity of the absolute calibration is confirmed with the data of standard stars in the LMC \citep{tanabe08} processed 
%with the latest imaging toolkit with the new flat data.
%Furthermore, by reducing data of standard stars in the LMC with the toolkit, 
%the validity of the flux calibration parameters given by \citet{tanabe08} is confirmed.

%We found there is a gentle decline to the direction of X-axis,
 %which corrected by fitting a quadratic curve to the systematic error (see bottom left panel of Figure.~\ref{fig_flat}).
%We have not found other effects to account for the systematic gradient yet,
% which is expected to some sort of scattering light in the optics.

\section{Data Reduction Techniques beyond the Imaging Pipeline Procedures for the Advanced Analyses of Diffuse Sources  \label{IRM}}
\begin{figure}[!ht]
\begin{center}
\includegraphics[width=\hsize,angle=0]{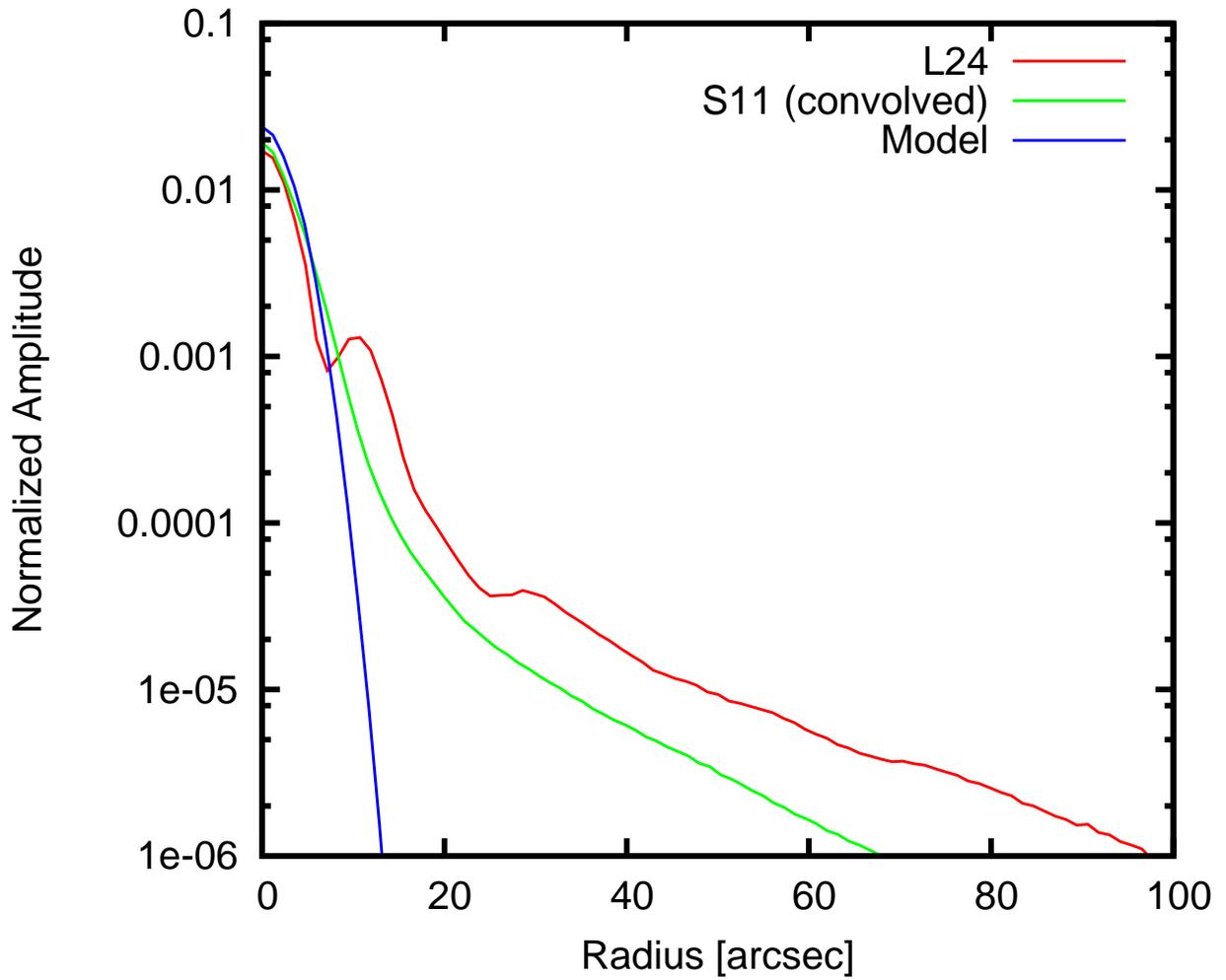}
\caption{Radial profiles of the PSF at L24 (red line),
the PSF at S11 (green line) convolved with a Gaussian to adjust the FWHM to 6\arcsec.7,
and the PSF obtained from the deconvolution-convolution process (equation (\ref{eqconvk})) for all the four bands (blue line).}
\label{radconv}
\end{center}
\end{figure}

\begin{figure}[!ht]
\begin{center}
  \begin{tabular}{cc}
      \resizebox{45mm}{!}{\includegraphics[angle=0]{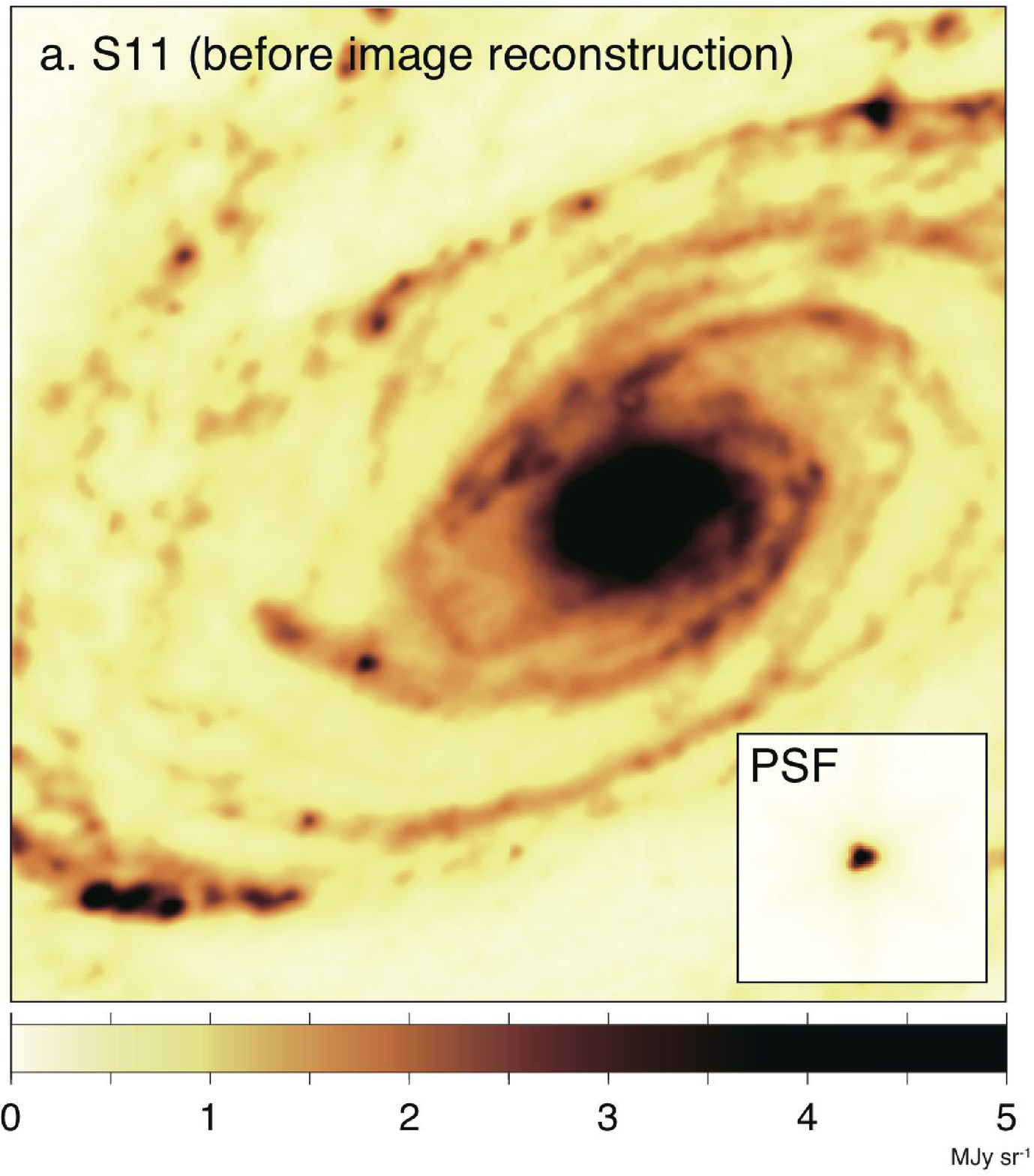}} &
      \resizebox{45mm}{!}{\includegraphics[angle=0]{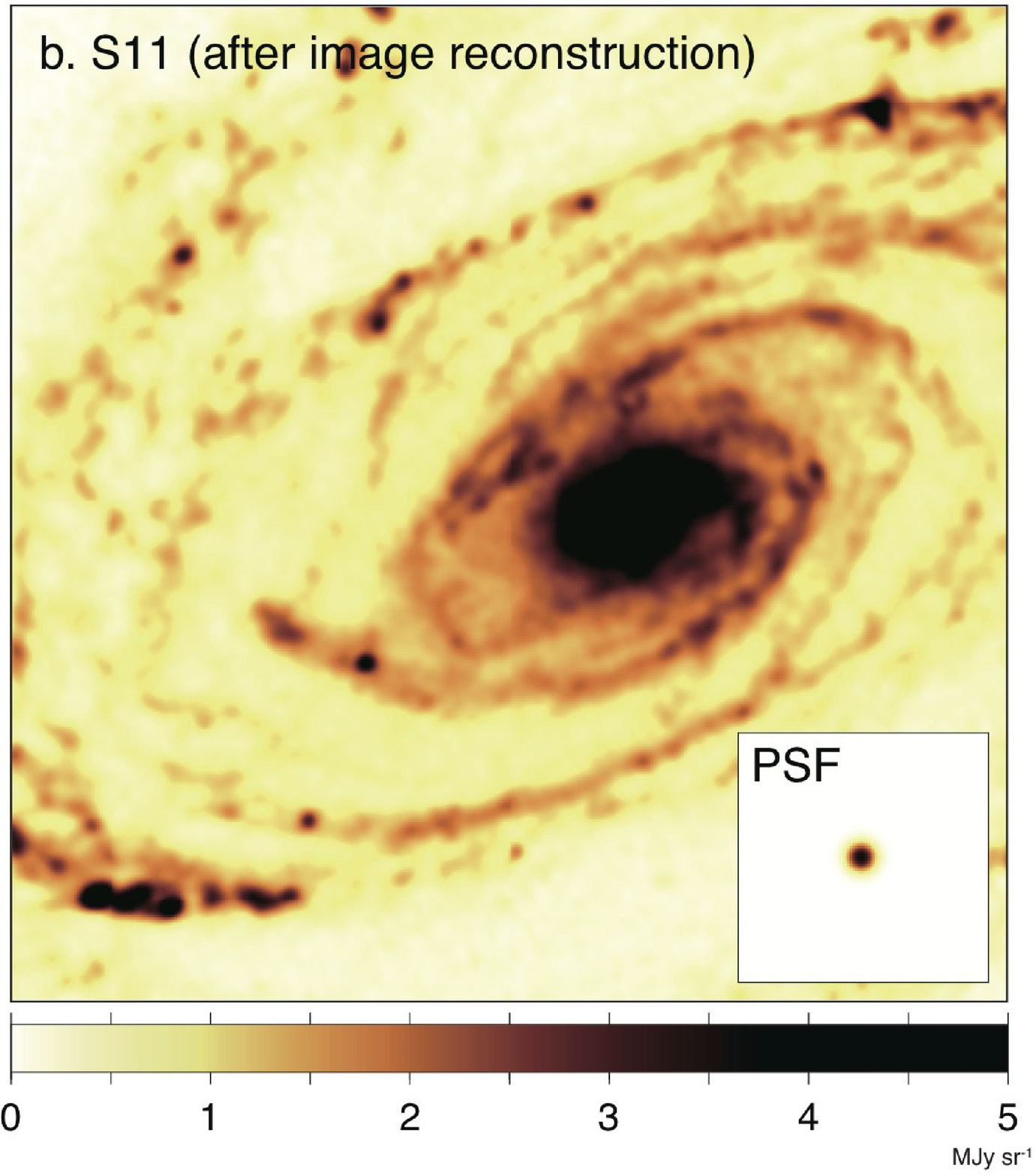}} \\
      \resizebox{45mm}{!}{\includegraphics[angle=0]{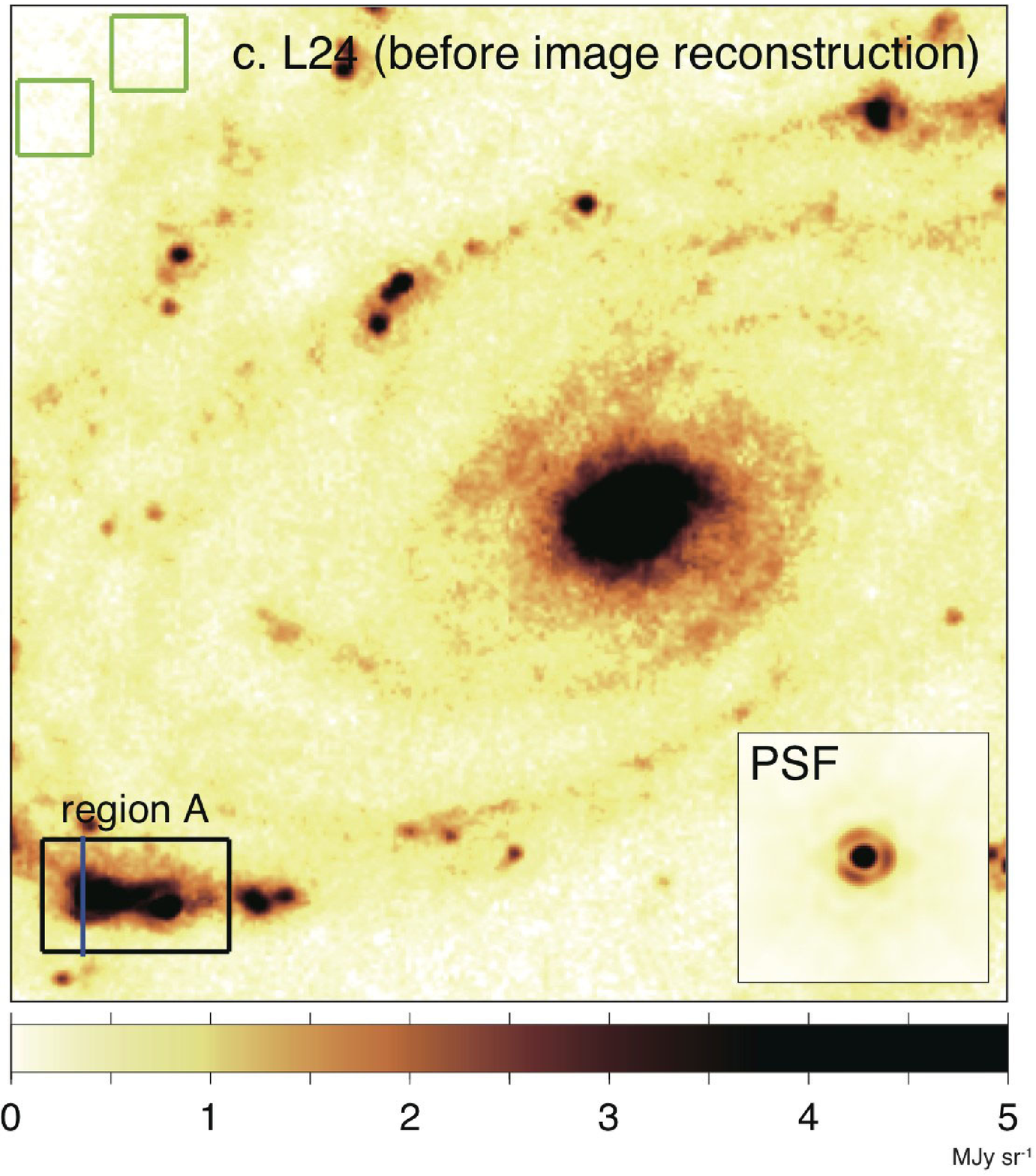}} &
      \resizebox{45mm}{!}{\includegraphics[angle=0]{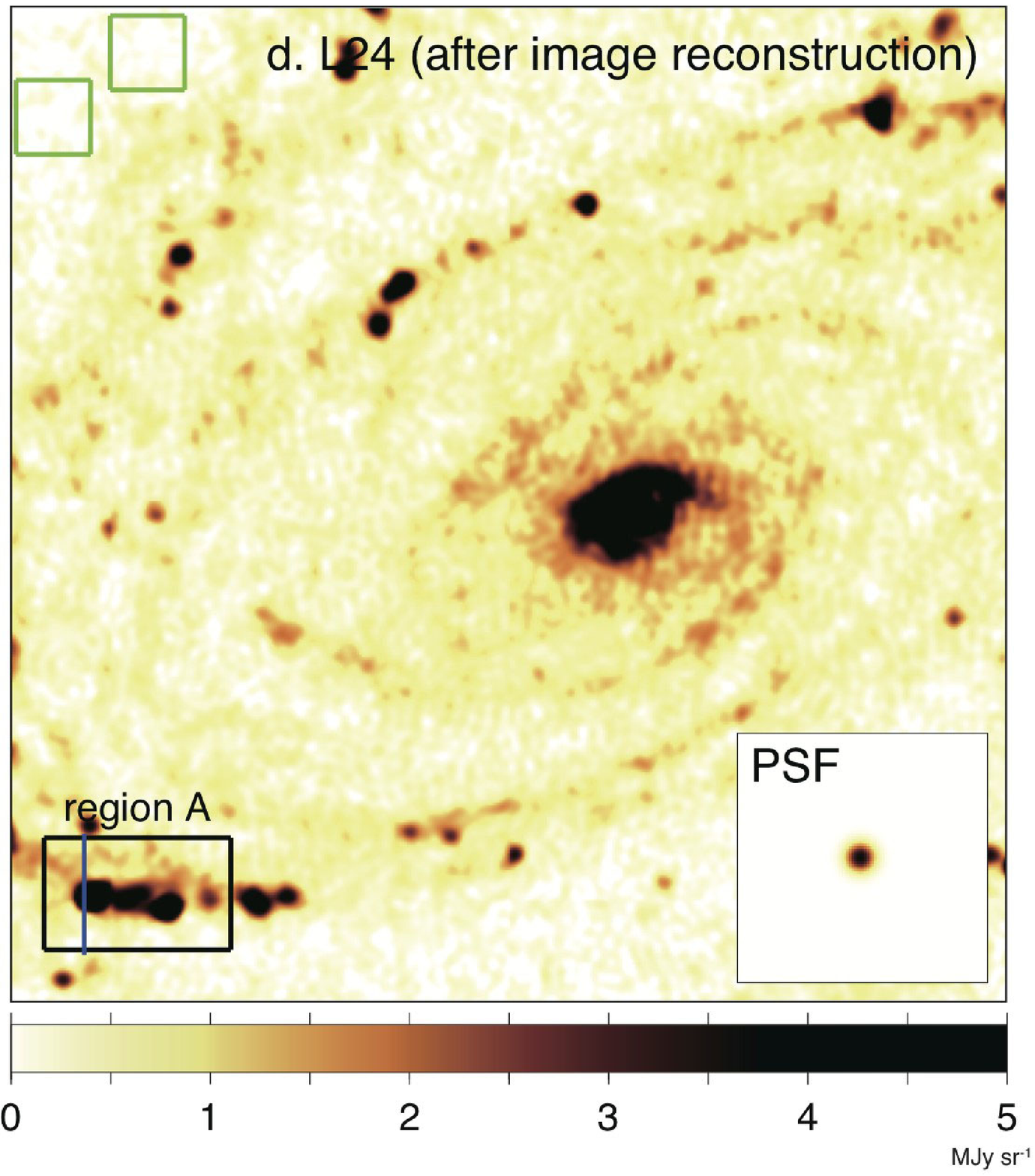}} \\
      \resizebox{45mm}{!}{\includegraphics[angle=0]{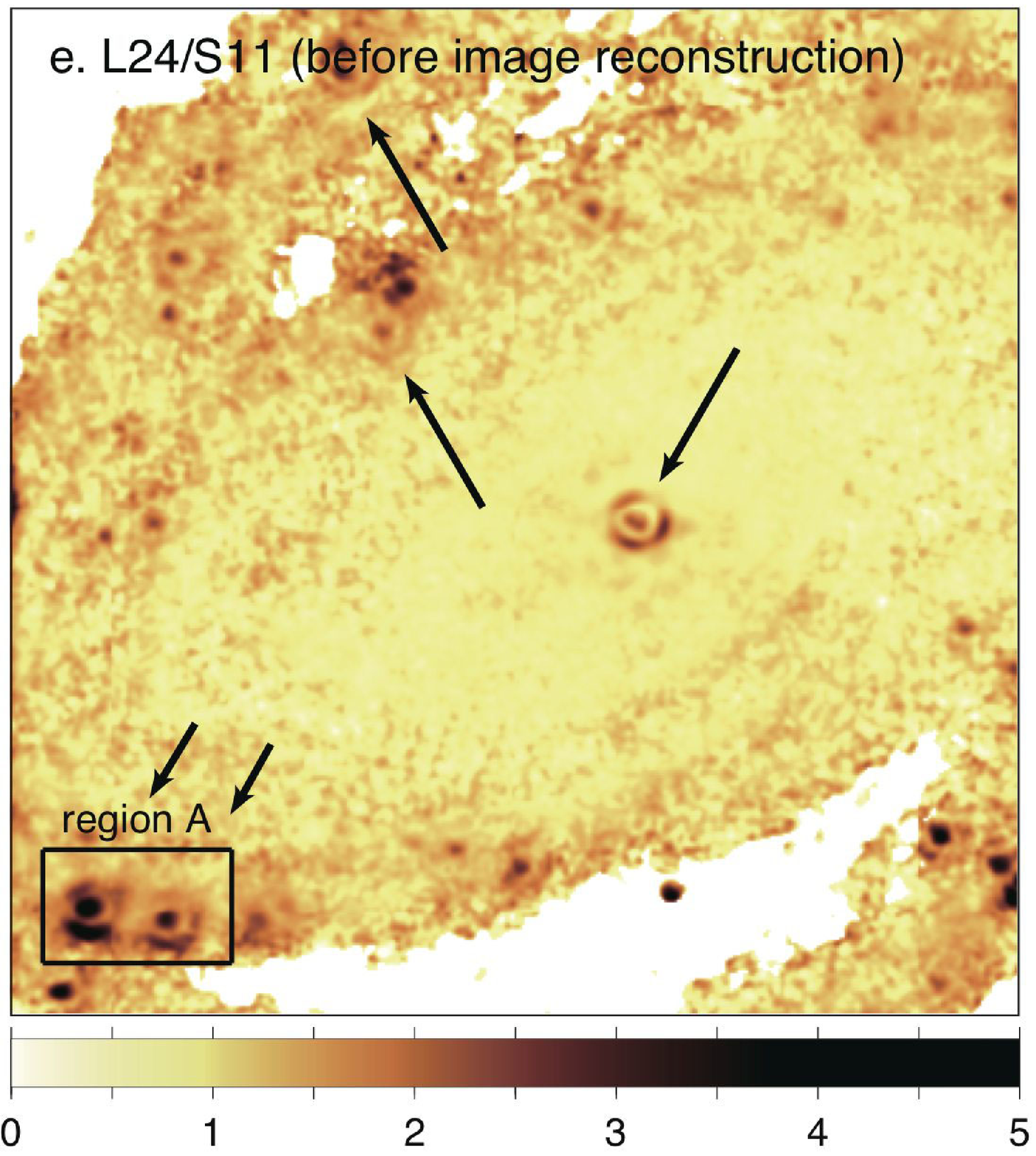}} &
      \resizebox{45mm}{!}{\includegraphics[angle=0]{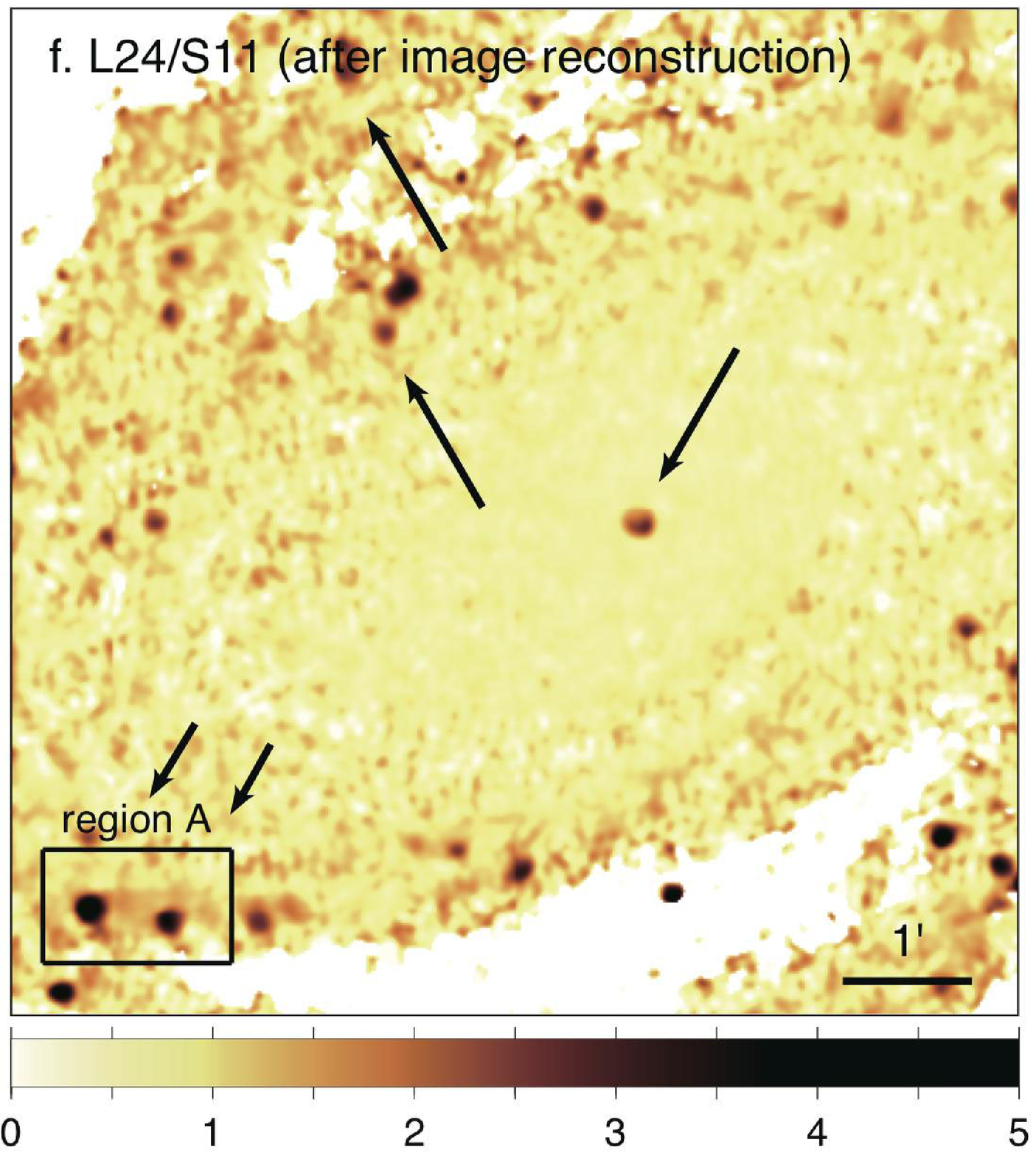}} \\
    \end{tabular}
\caption{
\small
Example of the image reconstruction process of the IRC mid-infrared data investigated in the present study (M81: pointing ID 5020062.1 and 5020063.1).
(a) Artifact-subtracted image of S11, 
(b) S11 image reconstructed with equation (\ref{eqconv}),
(c) artifact-subtracted image of L24,
(d) L24 image reconstructed with equation (\ref{eqconv}),
(e) the L24/S11 color map produced from the L24 image (c) and the S11 image (a) 
convolved with a simple Gaussian to adjust the FWHM to 6\arcsec.7 (green line in Figure~\ref{radconv}), 
and (f) the L24/S11 color map produced from the reconstructed images (b) and (d).
Arrows indicate bright \ion{H}{2} regions and the galactic center of M81, where ring-like artifacts appear in (c) and (e).
After the image reconstruction procedure, the ring-like artifacts seen in (c) and (e) disappear in (d) and (f). 
\textcolor{red}{The blue line in region A overlaid on (c) and (d) indicates the position of the surface brightness profiles shown in Figure~\ref{fig15} and the green boxes indicate the regions used to estimate the general fluctuation level (see \S~3).}
\label{rec1}}
\end{center}
\end{figure}

%As described in \S 1, 
A significant fraction of IRC observations in the cold phase were made with the S7, S11, L15, and L24 bands,
\textcolor{red}{which provide us useful information to characterize the emission properties of the interstellar matter (e.g., \citealt{onaka09,sakon07,onaka10}).
However, the differences in the PSF sizes as well as in the distinct patterns in a large scale, as shown in Table 5 and Figure 11, hamper us from directly making color analyses among those bands especially for extended objects such as ISM structures of nearby galaxies.}
%where emission from the interstellar matter dominates (e.g., \citealt{onaka10}).
%The variations in the relative intensities among the mid-infrared bands thus provide useful information on the study of the interstellar matter.
%%It is valuable to develop a method to compare the four-band images. 
%However, the FWHM of the PSFs differs significantly among the bands (Table~\ref{fwhm}).
%The PSFs also show distinct patterns in a large scale depending on the band (Figure~\ref{fig.psf}),
%which have to be taken into account.
%Therefore, direct comparison between the four bands and with numerous archival data is difficult especially for extended objects such as nearby galaxies.
Figure~\ref{radconv} compares the radial profile of the PSF at L24 
and 
\textcolor{red}{that} at S11 convolved with a Gaussian to have the same FWHM as the L24 PSF (6\arcsec.7, see Table~\ref{fwhm}). 
It shows that 
\textcolor{red}{the radial profile of} the adjusted PSF of S11 
is different from 
\textcolor{red}{that of} L24, 
especially at larger radii 
\textcolor{red}{beyond} $10\arcsec$.
%The L24/S11 flux ratio map of the nearby galaxy M81 made from this simple convolution adjustment of the S11 image
% shows ring-like artifacts around HII regions and the galactic center (Figure~\ref{rec1}e, indicated by the arrows).
\textcolor{red}{In Figure~\ref{rec1} we exemplify the case of imaging data of the nearby galaxy M81.}
Ring-like artifacts appear around \ion{H}{2} regions and 
the 
\textcolor{red}{nucleus of the galaxy}
in the L24/S11 flux ratio map of M81 made from this simple convolution of the S11 image (Figure~\ref{rec1}e, indicated by the arrows). 
The ring-like structures come from the diffraction pattern at L24,
\textcolor{red}{which should be corrected for by making use of the proper PSF pattern.}
To make comparative studies with the four bands it is necessary to adjust the PSF pattern properly.
We investigate a method to convolve the deconvolved image with an appropriate kernel.
%Even if the images of different bands are convolved to have the same FWHMs, 
%the output images are not simply compared because of the distinct patterns.
%To allow for accurate comparisons between wavebands, it is necessary to convolve all images to the same (generally the lowest) resolution.
%However, the PSFs of AKARI / IRC have complex structures which are differentiated with each other (Figure.~\ref{fig.psf}).

Following \citet{gordon08}, we define the convolution kernel $K(x,y)$ as
%We have developed an image reconstruction method using the PSF models based on the convolution kernels developed by \citet{gordon08}, described as follows;
\begin{eqnarray}
K(x,y) = {\rm DFT^{-1}} \bigg\{\frac{{\rm DFT[PSF_2}(x,y)]}{{\rm DFT[PSF_{1}}(x,y)]}  \bigg\}, 
\label{eqconvk} 
\end{eqnarray}
where DFT indicates the discrete Fourier transform operation and ${\rm DFT^{-1}}$ is the reverse operation.
${\rm PSF_1}$ is the PSF of the input image and ${\rm PSF_2}$ is a simple PSF to which all band images will be adjusted.
%\textcolor{red}{The X-Y positions of the PSFs is adjusted to the central pixel where the flux is the maximum.} 
%the desired output model PSF.
The output image $I_{\rm out}(X,Y)$ is then obtained by 
\begin{eqnarray}
I_{\rm out}(X,Y) = {\rm DFT^{-1}}\bigg\{ {\rm DFT}[K(x,y)] \cdot {\rm DFT}[I_{\rm in}(x,y)] \bigg\}.
\label{eqconv}  
\end{eqnarray} 
\textcolor{red}{In these operations, the sampling scale for $I_{\rm in}(X,Y)$, $I_{\rm out}(X,Y)$ 
and $K(x,y)$ is set as a half of the intrinsic pixel scale.}
%where $\times$ indicates the convolution operator.
\textcolor{red}{The} Gaussian 
\textcolor{red}{profile} with the FWHM of 7\arcsec 
\textcolor{red}{is employed for ${\rm PSF_2}$}.
The radial profile is shown by the blue line in Figure~\ref{radconv}.
%which is larger than that of the input image 
%to attenuate higher frequency noise in the images.
%It seems to be wasteful to blur the images with higher resolutions. 
The convolution with a Gaussian suppresses spurious features 
\textcolor{red}{found in} a simple deconvolution operation 
without losing spatial information significantly.

\begin{figure}[!ht]
\begin{center}
\includegraphics[width=\hsize,angle=0]{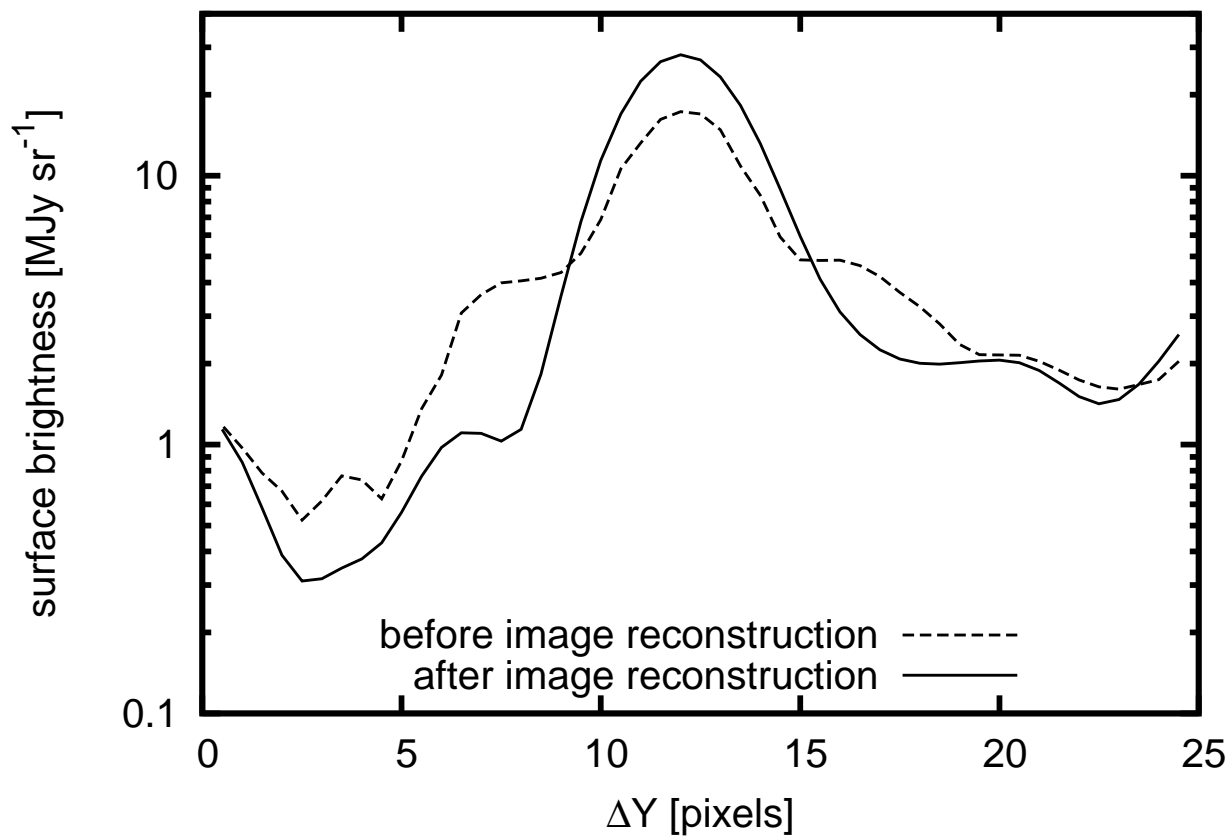}
\caption{\textcolor{red}{Surface brightness profiles at L24 along the line shown in Figures~\ref{rec1}c and d.
The dashed line represents the profile before image reconstruction (Figure~\ref{rec1}c); the solid line shows that after image reconstruction (Figure~\ref{rec1}d). The pixel scale is 2\arcsec.38, which corresponds to about $42$ pc assuming the distance of 3.6 Mpc to M81. A bright HII region is located at $\Delta Y = 12.5$ pixels and the ring-like pattern appears at $\Delta Y \sim 7.5$ pixels and $\sim 17.5$ pixels.} \label{fig15}}
\end{center}
\end{figure}

The present reconstruction procedure provides images that can be used reliably for the study of faint extended emission.
%\textcolor{red}{As an example of the image reconstruction procedure, we choose imaging data of nearby galaxy M81 (pointing ID 5020062.1 and 5020063.1) and make the procedure to the data.}
Figure~\ref{rec1} shows how the reconstruction procedure 
\textcolor{red}{described above} improves the image quality taken with the IRC.
%\textcolor{red}{In the figure we exemplify the case of imaging data of nearby galaxy M81.}
%The finally obtained images of S11 and L24 (Figure~\ref{rec1}b and d) have slightly reduced spatial resolution from the deconvolved images
%because of the convolution processes 
%given by equation (\ref{eqconv}) at a small price of the spatial resolution. 
%As shown in the reconstructed L24/S11 color map (Figure~\ref{rec1}f), 
%the ring-like structures disappear 
%and the contrast of the L24/S11 flux ratio is clearly enhanced
%compared to the original color map (Figure~\ref{rec1}e).
\textcolor{red}{For example, some discrete \ion{H}{2} regions are seen as bright point-like spots over the relatively faint diffuse emission from the ISM in the M81 in the L24 image in Figure~\ref{rec1}c.
%As shown in the same figure, diffuse emission is also detected in the same image.
%When performing the measurement 
Due to the extended component of the PSF, especially to the diffraction ring structure of the PSF at L24, the diffuse emission around such  bright sources are overlapped by the extended artifacts that come from the bright sources.
Figure 15 shows the surface brightness profiles along a certain selected line defined in both images of region A before and after processing the present image reconstruction (see Figures 14c and 14d. respectively).
%
%When measuring the surface brightness of a certain area of the diffuse structures 
%in region A, 
%the typical flux fluctuation level of these diffuse structures is approximately 0.7 MJy\,${\rm sr^{-1}}$.
%
A typical 
%fluctuation 
level of the diffuse emission in region A
is about 0.7 MJy\,${\rm sr^{-1}}$ if the area that is not affected
by bright sources.
On the other hand, 
%the ring-like patterns produced by a bright infrared point-like source centered at $\Delta Y = 12.5$ 
a bright infrared point-like source centered at $\Delta Y = 12.5$ with a flux density of 0.02 Jy at L24, for example, produces the ring-like pattern
%\ion{H}{2} region with a flux density of 0.2 Jy at L24 
%($\Delta Y = 12.5$ pixels in Figure~\ref{fig15})
%bright infrared point-like source centered at
%$\Delta Y = 12.5$ 
%with a flux density of 0.02 Jy at L24, for example, 
that makes a false contribution to the surface brightness as bright as 2 MJy\,${\rm sr^{-1}}$ 
in the profile ranges of $\Delta Y$=7-9 and $\Delta Y$=16-18 (see Figure 15).
%within the 
%%distance of 12\arcsec,
%profile ranges in between $\Delta Y =$ 7 and 18,
% which corresponds to 
%the regions within
%200 pc 
%from the peak position of the pointt source
%assuming the distance of 3.6 Mpc to M81.
In this case, about one third of the whole pixels in region A
%have signal ratios between the emission from diffuse sources and 
are almost comparably affected by 
the extended PSF components from the bright sources,
%relative to the emission from the diffuse sources in concern.
 which makes it difficult to estimate the emission from the diffuse sources in concern.
%of less than one, where it is difficult to estimate the fluxes of the diffuse sources accurately.
These artifacts are 
properly removed by the reconstruction procedure, 
as shown in Figures~\ref{rec1}d and \ref{fig15},
which offers significant 
%improvement in measuring the accurate surface brightness 
\textcolor{red}{improvement in accurate photometry of the surface brightness}
of diffuse sources located 
%at even closer region
\textcolor{red}{even at a closer region} 
to a point source.
%
%For a point source fainter than 0.1 Jy at L24, the surface brightness of the diffuse emission located at $>$ 5 pixels from the point source, which corresponds to the linear scale of 200 pc assuming the distance of 3.6 Mpc to M81, is obtained with an accuracy of 0.2 MJy\,${\rm sr^{-1}}$, which is roughly the general fluctuation level of the surface brightness of diffuse ISM structures found in regions without being affected by point sources (e.g., shown with green boxes in Figures 14c and 14d).
%
After the present image reconstruction,
the surface brightness of the diffuse emission can be measured with an
accuracy of 0.2 MJy\,${\rm sr^{-1}}$ at L24 
if there are no sources brighter than 0.1 Jy within 5 pixels from
the region in question. 
We note that the present methods do not take account of the changes in PSF shapes within the FOV. 
%and 
Thus the accuracy of the present image reconstruction may be
slightly degraded
for diffuse sources around a point source located in the corner of the
FOV.
%in the surface brightness of diffuse sources around a point source located in the corner of the FOV achieved via this reconstruction technique may be slightly degraded.
%
Figures~\ref{rec1}e and f indicate that the image reconstruction significantly improves the quality of the comparative study of four-band images.}
As shown in the reconstructed L24/S11 color map (Figure~\ref{rec1}f), 
the ring-like structures disappear 
and the contrast of the L24/S11 flux ratio is clearly enhanced
compared to the original color map (Figure~\ref{rec1}e).

\section{Summary \label{sum}}
Mid-infrared images taken with the IRC onboard {\it AKARI} suffer artifacts and extended PSFs.
The artifacts are investigated at four IRC bands, whose contributions to the total input signal are measured to be about $< 1 \%$, $1-3 \%$, $7 \%$, and $16 \%$ at S7, S11, L15, and L24, respectively.
These artifacts affect the flat-fielding reference data and new flat data are derived by removing the effects of the artifacts. 
The PSFs and the aperture correction factors for point and diffuse sources are obtained at S7, S11, L15, and L24.
%at S7, S11, L15, and L24 are produced for accurate photometry of both point and diffuse sources.
To adjust the spatial resolution at different bands, 
an image reconstruction method is developed.
The present investigation enables us to carry out the comparative study of four-band mid-infrared images taken with the IRC.
\textcolor{red}{The artifacts and extended PSFs of the IRC data have common characteristics inherent in mid-infrared instruments.  
The present analysis will be useful not only for the analysis of IRC MIR data,  
but also for the design, ground tests, 
on-orbit calibration plans, and data analyses of mid-infrared instruments of future facilities.
The image reconstruction developed here can also be applied to data taken with other instruments.}

The MIR-L ghost patterns, the new flat data of L15 and L24, and the HDR PSFs
of S7, S11, L15, and L24 are available at
http://www.ir.isas.jaxa.jp/AKARI/Observation/.
The new flat data are included in the latest version of the imaging toolkit (ver. 20110225).
The validity of the absolute calibration is confirmed with the data of standard stars in the LMC \citep{tanabe08} processed 
with the latest imaging toolkit with the new flat data.

\acknowledgments

This work is based on observations with {\it AKARI}, a JAXA project
with the participation of ESA.  The authors thank all the members of
the {\it AKARI} project.  
This work utilizes the IRC data taken during the performance
verification phase.  They are grateful to the IRC team members for their continuous
encouragement and useful comments.
We thank N. Fujishiro for making ray-tracing calculations of the
artifacts.
D. K. is supported by a Grant-in-Aid from the Japan Society for the
Promotion of Science.
\textcolor{red}{This work is supported in part by a Grant-in-Aid from the
Japan Society of Promotion of Science (JSPS).}

\newpage
\begin{table}[!ht]
\begin{center}
\caption{List of the datasets used for PSFs,artifacts and flat-fielding investigation\label{tab4}}
\vspace{0.5truemm}
\tiny
\begin{tabular}{lcccc}
\hline Observation ID  & AOT\tablenotemark{a} &  Date & Object  & Note \\ \hline
MIR-S channel& & & & \\
4080002.1 & IRC03 a;L &2007 Mar 05  & CRL618        & PSF \& SCL\\
4080019.1 & IRC03 a;L & 2007 Mar 05  &CRL618        & PSF      \\
4080007.1 & IRC03 a;L &2006 Sep 28  & Red Rectangle & PSF \& SCL \\
4080006.1 & IRC03 a;L &2006 Sep 28  & Red Rectangle & PSF \\
1800623.1 & IRC02 a;L &2007 May 10  & HIP 28909     & PSF   \\
5020008.1 & IRC03 a;L &2006 Apr 22  & HIP 28909     & PSF   \\
5124007.1 & IRC03 a;L &2006 Jun 24  & KF01T4        & PSF   \\
5124019.1 & IRC03 a;L &2006 Aug 03  & KF03T2        & PSF   \\
1400229.1 & IRC04 b;Ns &2007 Feb 06 & CAR 057TO005  &SCL \\
1401033.1 & IRC04 b;Ns &2007 Apr 06 & CRU 032 S003  &SCL \\
5020048.1 & IRC04 a;Ns &2006 Apr 29 & NGC6543 &SCL \\
 \hline
%\end{tabular}
%\end{center}
%\end{table}
%
%\newpage
%\begin{table}[!ht]
%\begin{center}
%\caption{List of the data sets for MIR-L PSFs and artifacts \label{tab5}}
%\vspace{0.5truemm} 
%\begin{tabular}{lcccc}\hline
%Observation ID & AOT & Date & Object & Note \\ \hline
MIR-L channel & & & & \\
1711365.1 & IRC02 a;L  & 2007 May 12  & IRAS 22396$-$4708 & PSF \& SCL  \\
4080019.1 & IRC03 a;L  & 2007 Mar 5  & CRL618            & SCL \\
4080002.1 & IRC03 a;L  &2007 Mar 5  & CRL618        & PSF\& SCL  \\
5124105.1 & IRC03 a;L  &2007 Jul 12 & IRAS F06009$-$6636& PSF \& FLAT \\
5124089.1 & IRC03 a;L  &2007 Jun 19  & IRAS F06009$-$6636& PSF \& FLAT \\
5124090.1 & IRC03 a;L  &2007 Jun 22  & IRAS F06009$-$6636& PSF \& FLAT \\
5124091.1 & IRC03 a;L  &2007 Jun 22  & IRAS F06009$-$6636& PSF \& FLAT  \\
5124099.1 & IRC03 a;L  &2007 Jul 4  & IRAS F06009$-$6636& PSF \& FLAT  \\
5124092.1 & IRC03 a;L  &2007 Jul 21  & IRAS F06009$-$6636& PSF \& FLAT \\
5124098.1 & IRC03 a;L  &2007 Jul 4  & IRAS F06009$-$6636& PSF \& FLAT  \\
5124097.1 & IRC03 a;L  &2007 Jul 1  & IRAS F06009$-$6636& PSF \& FLAT  \\
5124100.1 & IRC03 a;L  &2007 Jul 5  & IRAS F06009$-$6636& PSF \& FLAT  \\
5124083.1 & IRC03 a;L  &2007 Jun 1  & IRAS F06009$-$6636& FLAT  \\
5124081.1 & IRC03 a;L  &2007 Jun 2  & IRAS F06009$-$6636& FLAT  \\
5124082.1 & IRC03 a;L  &2007 Jun 2  & IRAS F06009$-$6636& FLAT  \\
5124084.1 & IRC03 a;L  &2007 Jun 7  & IRAS F06009$-$6636& FLAT  \\
5124085.1 & IRC03 a;L  &2007 Jun 12 & IRAS F06009$-$6636& FLAT  \\
5124087.1 & IRC03 a;L  &2007 Jun 13 & IRAS F06009$-$6636& FLAT  \\
5124086.1 & IRC03 a;L  &2007 Jun 14  & IRAS F06009$-$6636& FLAT  \\
5124088.1 & IRC03 a;L  &2007 Jun 18  & IRAS F06009$-$6636& FLAT  \\
5124094.1 & IRC03 a;L  &2007 Jun 26 & IRAS F06009$-$6636& FLAT  \\
5124095.1 & IRC03 a;L  &2007 Jun 27& IRAS F06009$-$6636& FLAT  \\
5124093.1 & IRC03 a;L  &2007 Jun 29 & IRAS F06009$-$6636& FLAT  \\
5124096.1 & IRC03 a;L  &2007 Jun 30 & IRAS F06009$-$6636& FLAT  \\
5124101.1 & IRC03 a;L  &2007 Jul 8& IRAS F06009$-$6636& FLAT  \\
5124102.1 & IRC03 a;L  &2007 Jul 8& IRAS F06009$-$6636& FLAT  \\
5124103.1 & IRC03 a;L  &2007 Jul 9& IRAS F06009$-$6636& FLAT  \\
5124104.1 & IRC03 a;L  &2007 Jul 12& IRAS F06009$-$6636& FLAT  \\
 \hline
\end{tabular}
\tablenotetext{a}{Astronomical Observation Template for the IRC and FIS observations. See 
ASTRO-F Observer's Manual for details of the parameters
({\tt http://www.ir.isas.jaxa.jp/AKARI/Observation/ObsMan/}).}
\tablenotetext{b}{PSF indicates the data that were used to derive the PSF.  SCL indicates
those used to estimate the scattered light contribution (artifacts).
FLAT indicates those used to estimate the reference data for flat-fielding at L15 and L24.}
\end{center}
\end{table}

\clearpage 
\begin{table}[!ht]
\begin{center}
  \caption{Relative intensity of the MIR-S artifact}
%\vspace{0.5truemm} 
\begin{tabular}{lcc}
\tableline
Object   & S7 & S11 \\ 
\tableline
A-type star\tablenotemark{a}    & $0.0065 \pm 0.0008$ & $0.016 \pm 0.002$  \\
K-type star\tablenotemark{b}    & $0.0073 \pm 0.0006$ & $0.024 \pm 0.004$ \\
Zodiacle light\tablenotemark{c} & $<0.0027$ & $0.023 \pm 0.008$\\
\tableline
\label{aa}
\end{tabular}
\tablenotetext{a}{HD 28909 (A0 star)}
\tablenotetext{b}{KF01T4 (K1.5III star)}
\tablenotetext{c}{Calculated from the slit image.}
\end{center}
\end{table}

\newpage
\begin{table}[hhhh]
\begin{center}
  \caption{Parameters for MIR-L L15 and L24 artifact movements}
  \label{tab:03}
%\begin{center}
\vspace{0.8truemm} 
\begin{tabular}{lcc}\tableline
Artifact      &  $a$  [pixels]\tablenotemark{a} &  $b$ [pixels]\tablenotemark{a}    \\ \tableline
    A      & $0.364 \times (1\pm 0.051)$  &  $0.194\times (1\pm 0.012)$   \\ 
    B and C & $-1.426\times (1\pm 0.0049) $  & $-1.272\times (1\pm 0.0028) $        \\ \tableline
  \end{tabular}
\tablenotetext{a}{The fit parameters in equation (\ref{eqart}).}
\end{center}
\end{table}

\begin{table}[!ht]
\begin{center}
  \caption{Properties of the artifacts at L15 and L24}
  \label{tab:01}
\vspace{0.8truemm}
%\begin{table}[hhh]
  \begin{tabular}{lccc}\tableline
Artifact   &  Total signal to & Typical radius  & Surface brightness per pixel \\
  & the object signal &    [pixels] & relative to  the  object signal  \\ \tableline
L15  & & & \\
A & $2.5\times 10^{-2}$  & $~100$  & $1 \times 10^{-6}  $  \\ 
     B         & $1.8\times 10^{-2}$   & $30$       & $2 \times 10^{-6}  $      \\ 
     C         & $3.0\times 10^{-2}$   &  $~150$    &     $6 \times 10^{-7}  $      \\ \tableline
L24 & & & \\
    A & $6.8\times 10^{-2}$  & $~100$  & $2.4 \times 10^{-6}  $  \\ 
    B          & $4.9\times 10^{-2}$   & $30$       &     $1.8 \times 10^{-5}  $      \\ 
    C          & $4.9\times 10^{-2}$   &  $~150$    &     $1 \times 10^{-6}  $      \\ \tableline
  \end{tabular}
\end{center}
\end{table}

%%\begin{table}[hhhh]
%%\begin{center}
%  \caption{Parameters for MIR-L L15 and L24 artifact movements. }
%  \label{tab:03}
%%\begin{center}
%\vspace{0.5truemm} 
%\begin{tabular}{lcc}\tableline
%Artifact      &  $a$  [pixels]\tablenotemark{a} &  $b$ [pixels]\tablenotemark{a}    \\ \tableline
%    A      & $0.385 \times (1\pm 0.051)$  &  $0.194\times (1\pm 0.012)$   \\ 
%    B \& C & $-1.426\times (1\pm 0.0049) $  & $-1.272\times (1\pm 0.0028) $        \\ \tableline
%  \end{tabular}
%\tablenotetext{a}{The fit parameters in Equation (\ref{eqart}).}
%\end{center}
%\end{table}

\clearpage 
\begin{table}[!ht]
\begin{center}
  \caption{FWHMs of the PSFs}
  \label{fwhm}
\vspace{0.8truemm}
%\vspace{0.5truemm} 
\begin{tabular}{lc}
\tableline
 IRC band    & FWHM [arcsec]    \\ 
\tableline
S7  & 2.9 \\
S11 & 3.3 \\
L15 & 4.6 \\
L24 & 6.7\\ 
\tableline
\end{tabular}
\end{center}
\end{table}

\begin{table}[!ht]
\begin{center}
  \caption{Best-fit coefficients of the aperture correction factors}
  \label{tab:a04}
\vspace{0.8truemm}
%\vspace{0.5truemm} 
\begin{tabular}{lccc}
\tableline
 IRC band    & $A$\tablenotemark{a}  & $B$\tablenotemark{a}  & $C$\tablenotemark{a}     \\ 
\tableline
S7  & 0.776 & 0.371 & 0.906 \\
S11 & 0.804 & 0.428 & 0.926 \\
L15 & 1.671  & 0.417 & 0.843 \\
L24 & 1.764  & 0.396 & 0.817 \\ 
\tableline
\end{tabular}
\tablenotetext{a}{The fit parameters in equation (\ref{eqAC}).}
\end{center}
\end{table}

\newpage

\newpage

\newpage

\newpage

\newpage

\clearpage

\clearpage

\clearpage

\clearpage

\clearpage

\clearpage

\clearpage

\clearpage

%\begin{table}[!ht]
%\begin{center}
%  \caption{The obtained diffuse calibration factors.}
%  \label{tab:a06}} 
%\begin{tabular}{lc}
%\tableline
% IRC band & Diffuse Calibration Factor \\ 
%\tableline
%S7  & 0.91 \\
%S11 & 0.93 \\
%L15 & 0.84 \\
%L24 & 0.82 \\ 
%\tableline
%\end{tabular}
%\end{center}
%\end{table}

\end{document}